\documentclass[a4paper,12pt]{article}
\usepackage{graphicx,bbm}
\usepackage{amsmath,amsfonts,amssymb}
\usepackage{fleqn} 
\usepackage{mathrsfs}
\usepackage[center,footnotesize,hang]{subfigure}
\usepackage{epsfig}\parskip 5pt plus 1pt
\usepackage[footnotesize]{caption}

\newcommand{\CenterEps}[2][1]{\ensuremath{\vcenter{\hbox{\includegraphics[scale=#1]{#2.eps}}}}}

\def\imag{\mathrm{i}}

\def\beq{\begin{equation}}
\def\eeq{\end{equation}}
\def\bea{\begin{eqnarray}}
\def\eea{\end{eqnarray}}

\newcommand{\gsim}{\lower.7ex\hbox{$\;\stackrel{\textstyle>}{\sim}\;$}}
\newcommand{\lsim}{\lower.7ex\hbox{$\;\stackrel{\textstyle<}{\sim}\;$}}

\def\<{\left\langle}
\def\>{\right\rangle}

\allowdisplaybreaks[2]
\begin{document}

\bibliographystyle{OurBibTeX}

\begin{titlepage}

\vspace*{-15mm}
\begin{flushright}
FTUAM 06-17\\ 
IFT-UAM/CSIC 06-54\\
LPT-ORSAY-06-69\\
hep-ph/0611232
\end{flushright}
\vspace*{9mm}

\begin{center}
{\Large\textbf{
Towards constraints on the SUSY seesaw   \\[2mm] 
from flavour-dependent leptogenesis 
}}
\\[8mm]
\vspace*{0.10cm}

{\large
S.~Antusch$^{a}$ 
and A.~M.~Teixeira$^{a,b}$}\\[0.4cm]

$^{a}$ 
{\textit{Departamento de F\'{\i }sica Te\'{o}rica C-XI and }}\\

{\textit{Instituto de F\'{\i }sica Te\'{o}rica C-XVI, \\
Universidad Aut\'{o}noma de Madrid,
Cantoblanco, E-28049 Madrid, Spain}}\\[0pt]

\vspace*{0.2cm} 
$^{b}$ 
{\textit{Laboratoire de Physique Th\'eorique}}\\

{\textit{Universit\'e de Paris-Sud XI, \\
B\^atiment 201, F-91405 Orsay Cedex, France}}\\[0pt]

\end{center}

\vspace*{0.50cm}

\begin{abstract}
\noindent 
We systematically investigate constraints on the parameters of the
supersymmetric type-I seesaw mechanism from the requirement of successful
thermal leptogenesis in the presence of upper bounds on the reheat
temperature $T_\mathrm{RH}$ of the early Universe.
To this end, we solve the flavour-dependent Boltzmann equations in the MSSM,
extended to include reheating. 
With conservative bounds on $T_\mathrm{RH}$, leading to mildly constrained 
scenarios for thermal leptogenesis, compatibility with observation 
can be obtained for extensive new regions of the parameter space, due to 
flavour-dependent effects. 
On the other hand, focusing on (normal) hierarchical light and heavy neutrinos, 
the hypothesis that there is no CP violation associated with the right-handed neutrino sector, and that leptogenesis exclusively arises from
the CP-violating phases of the $U_\text{MNS}$ 
matrix, is only marginally consistent. Taking into account stricter
bounds on $T_\mathrm{RH}$ further suggests that (additional) sources of CP
violation must arise from the right-handed neutrino sector, further implying
stronger constraints for the right-handed neutrino parameters.
\end{abstract}
\end{titlepage}
\newpage
\setcounter{footnote}{0}

\section{Introduction}

One of the most appealing mechanisms to generate the observed baryon
asymmetry of the Universe (BAU), $n_\mathrm{B} /n_\gamma \,\approx\,
(6.10\,\pm\,0.21)\,\times\,10^{-10}$~\cite{Spergel:2006hy}, 
is that of baryogenesis via leptogenesis. 
Thermal leptogenesis is an attractive and minimal mechanism, in which
a lepton asymmetry is dynamically generated, and then 
converted into a baryon asymmetry
due to $(B+L)$-violating sphaleron interactions~\cite{Kuzmin:1985mm}.
The latter exist both in the Standard Model (SM) and in its minimal
supersymmetric (SUSY) extension, the MSSM.

The seesaw mechanism~\cite{seesaw:I,seesaw:II} provides an elegant
explanation for the observed smallness of the 
neutrino masses and, in addition, it 
offers the possibility of leptogenesis~\cite{FY}. In this case, the 
lepton asymmetry is generated
by the out-of-equilibrium decays of the same heavy right-handed
neutrinos which are responsible for the suppression of the light
neutrino masses.  
In spite of being one of the most simple frameworks where thermal
leptogenesis can be realised, the seesaw mechanism introduces a large
number of new parameters (in both its SM and MSSM versions). Although an
important amount of data has already been collected, many of the seesaw
parameters, namely those associated with the right-handed
neutrino sector, are experimentally unreachable.
As discussed by many authors, strong constraints on the
seesaw parameter space can be imposed from the requirement of
successful thermal leptogenesis~\cite{lept}. 
Typically, all these studies relied 
on the so-called flavour-independent (or one-flavour) approximation to
thermal leptogenesis. In the latter approximation,
the baryon asymmetry is calculated by
solving a Boltzmann equation for the abundance of
the lightest right-handed neutrino, and for the total lepton
asymmetry. Additionally, in the flavour-independent approximation,
the only relevant CP violation sources are those associated with the
right-handed neutrino sector (more concretely, the complex
$R$-matrix angles, when working in the so-called 
Casas-Ibarra parameterisation~\cite{Casas:2001sr}).

In recent years, the impact of flavour in thermal leptogenesis has
merited increasing attention~\cite{barbieri}-\cite{Branco:2006ce}. 
In fact, the one-flavour
approximation is only rigorously correct when the interactions mediated by 
the charged lepton Yukawa couplings are out of equilibrium. Below a given
temperature (e.g. $\mathcal{O}(10^{12}\,\text{GeV})$ in the SM), the 
tau Yukawa coupling comes into equilibrium (later followed by the couplings
of the muon and electron). 
Flavour effects are then physical and become manifest, not only at
the level of the generated CP asymmetries, but also regarding the
washout processes that destroy the asymmetries created for each
flavour. In the full computation, 
the asymmetries in each distinguishable flavour are
differently washed out, and appear with distinct weights in the 
final baryon asymmetry.

Flavour-dependent leptogenesis has recently been 
addressed in detail by several 
authors~\cite{davidsonetal}-\cite{Antusch:2006cw}. In particular, 
flavour-dependent effects in leptogenesis have been studied, 
and shown to be relevant, in the two
right-handed neutrino models~\cite{Abada:2006ea} as well as in classes
of neutrino mass models with three right-handed 
neutrinos~\cite{Antusch:2006cw}.
The quantum oscillations/correlations of the asymmetries in 
lepton flavour space have been included in Ref.~\cite{davidsonetal} 
and the treatment has also 
been generalised to the MSSM~\cite{Antusch:2006cw}. 

One interesting implication of the flavour-dependent
treatment is that in addition to the right-handed sector 
CP violating phases (that is, a complex $R$-matrix), 
low-energy CP violating sources, associated with the
light neutrino sector, also play a relevant role. 
Even in the absence of CP violation from the 
right-handed neutrino sector (which would lead to a zero baryon
asymmetry in the one-flavour approximation), a non-vanishing
baryon asymmetry can in principle be generated from the CP phases in
the Maki-Nakagawa-Sakata matrix, $U_\text{MNS}$. 
Strong connections between the low-energy CP phases of the 
$U_\text{MNS}$ matrix and CP violation for flavour-dependent 
leptogenesis can either emerge in classes
of neutrino mass models~\cite{Antusch:2006cw} or under the hypothesis
of no CP violation sources associated with the right-handed neutrino 
sector (real $R$) \cite{Blanchet:2006be,Pascoli:2006ie,Branco:2006ce}.
In addition, in the latter limit,
bounds on the masses of the light and heavy neutrinos and on the 
flavour-dependent decay asymmetries have been
derived~\cite{Branco:2006ce}.
The correlation of the baryon asymmetry with the effective Majorana
mass in neutrinoless double beta ($0\nu\beta\beta$) decays has also
been addressed~\cite{Pascoli:2006ie}.
Another appealing aspect of the flavour-dependent
treatment is that at least one of the CP sources, 
the Dirac CP-violating phase, is likely to be experimentally measured
(one of the Majorana phases could also in principle be measured - 
even though this represents a considerable
challenge~\cite{Pascoli:2002qm}, while the 
right-handed phases are experimentally unaccessible).

In the supersymmetric implementation of the seesaw mechanism, further
constraints on the seesaw parameter space can arise. 
These are particularly relevant in models of local SUSY (i.e. supergravity). 
First, one should consider cosmological
bounds on the reheat temperature ($T_\text{RH}$) after inflation, 
associated with the thermal production of gravitinos.
$T_\text{RH}$ 
has generally a strong impact on thermal leptogenesis, since the
production of right-handed neutrinos is suppressed if their mass
largely exceeds $T_\text{RH}$. 
In fact, due these bounds on the reheat temperature, viable thermal 
leptogenesis will impose strong constraints on the seesaw parameter 
space of locally supersymmetric models. 
Secondly, additional bounds on the SUSY seesaw parameter space arise
from low-energy observables, namely lepton flavour violating (LFV)
muon and tau decays such as 
$l_j \to l_i \, \gamma$ and $l_j \to 3 \, l_i$ ({\small$i<j$}), and   
charged lepton electric dipole moments. 
In studies of LFV, thermal leptogenesis in the  
flavour-independent approximation has been discussed 
in Ref.~\cite{LeptAndLFV}, and reheating effects have been explicitly
included in Ref.~\cite{Antusch:2006vw}.     
With a potential future observation of the sparticle spectrum, the
combined constraints on the SUSY seesaw parameters from leptogenesis
and LFV could lead to interesting information on the heavy
neutrino sector, which is otherwise unobservable at accelerators. 

In this study, our aim is to investigate the
constraints on the parameters of the type-I SUSY seesaw  
mechanism from the requirement of a successful flavour-dependent
thermal leptogenesis in the presence of upper bounds on the reheat
temperature of the early Universe.  
Previous studies of flavour-dependent thermal leptogenesis 
were conducted in the SM, 
and for a mass range of the lightest right-handed neutrino
where only the tau-flavour is in thermal equilibrium.
In the present 
analysis, we will work in the context of the MSSM 
extended by three right-handed neutrino
superfields. Moreover, for the temperatures we will consider, both 
tau- and muon-flavours are in thermal equilibrium, so that in fact all 
leptonic flavours must be treated separately.
We update the Boltzmann equations of Ref.~\cite{Giudice:2003jh} to include
flavour effects,  
and point out which regions of the seesaw parameter space 
generically enable optimal
efficiency and/or optimal decay asymmetries for leptogenesis
(focusing on the case of hierarchical light and heavy 
neutrino masses). We then discuss the  
differences between the flavour-independent approximation and the 
correct flavour-dependent treatment. 
We encounter interesting new regions of the seesaw parameter space, which are
now viable due to flavour-dependent effects. 
On the other hand, and
as we will discuss throughout this work, scenarios of leptogenesis solely
arising from the $U_\text {MNS}$ phases are quite difficult to accommodate,
and become increasingly disfavoured when stronger bounds on  
$T_\text{RH}$ are taken into account.

In the presence of strict bounds on $T_\text{RH}$, 
whether or not it is possible to generate the
observed baryon asymmetry exclusively from low-energy Dirac and/or
Majorana phases is still a question that deserves careful consideration.
Likewise, it is worth considering to which extent 
right-handed sector phases (other than those of the $U_\text{MNS}$) could
affect a potentially 
viable scenario of low-energy CP violating leptogenesis, and vice-versa. 

Our work is organised as follows. In Section~\ref{model:susyseesaw},
we briefly summarise the most relevant aspects of the type-I
SUSY seesaw mechanism. 
Section~\ref{Sec:FML} is devoted to the discussion of thermal
leptogenesis with reheating. In addition to estimating the baryon
asymmetry, we focus on the constraints on the reheat temperature
arising from the gravitino problem. A short summary of the limitations
and approximations of our computation is also included. In
Section~\ref{sec:constraints}, we finally discuss the constraints on
the seesaw parameters obtained from the requirement of successful
leptogenesis, namely constraints 
on the neutrino masses and on the $R$-matrix
mixing angles. Our conclusions are presented in Section~\ref{concl}.

\section{The seesaw mechanism and its parameters}\label{model:susyseesaw}

In what follows, we briefly introduce the most relevant features of 
neutrino mass generation via the seesaw mechanism. 
In the MSSM extended by three right-handed neutrino superfields,
the relevant terms in the superpotential to describe a type-I SUSY
seesaw are
\begin{equation}\label{W:Hl:def}
W\,=\,\hat N^c\,\lambda_\nu\,\hat L \, \hat H_2 \,+\,
\hat E^c\,\lambda_\ell\,\hat L \, \hat H_1 \,+\,
\frac{1}{2}\,\hat N^c\,m_M\,\hat N^c\,.
\end{equation}
In the above, $\hat N^c$ denotes the additional superfields containing 
the right-handed neutrinos $N_i$ and sneutrinos $\widetilde N_i$.  
The lepton Yukawa couplings $\lambda_{\ell,\nu}$ and the
Majorana mass $m_M$ are $3\times 3$ matrices in lepton flavour
space.
From now on, we will assume that we are in a basis where 
$\lambda_\ell$ and $m_M$ are diagonal.
After electroweak (EW) symmetry breaking, the charged lepton and 
Dirac neutrino mass matrices can be explicitly 
written as $m_\ell\,=\,\lambda_\ell\,\,v_1$, 
$m_D\,=\,\lambda_\nu\,v_2$, 
where $v_i$ are the vacuum expectation values (VEVs) of the neutral Higgs
scalars, with $v_{1(2)}= \,v\,\cos (\sin) \beta$ and $v=174$
GeV.

The $6\times 6$ neutrino mass matrix, whose eigenvalues are the
masses of the six Majorana neutrinos,   
is given by
\begin{equation}\label{seesaw:def}
M^\nu\,=\,\left(
\begin{array}{cc}
0 & \lambda^T_\nu\,v_2 \\
\lambda_\nu\,v_2 & m_M
\end{array} \right)\,. 
\end{equation}
In the seesaw limit, $m_{M_i}\,\gg\,v_2$, 
one obtains three light and three heavy
states, $\nu_i$ and $N_i$, respectively. 
Block-diagonalisation of the neutrino mass matrix of
Eq.~(\ref{seesaw:def}), leads (at lowest order in the
$(\lambda_\nu\,v_2/m_M)^n$ 
expansion) to the standard seesaw equation 
for the light neutrino mass matrix,
\begin{equation}\label{seesaw:light}
m_\nu\,=\, - v_2^2 \,\lambda_\nu^T \,m_M^{-1}\, \lambda_\nu  \,, 
\end{equation}
and to $m_{N}\,\simeq\,m_M$.
Since we are working in a basis where $m_M$ is diagonal, the heavy
eigenstates are then given by 
\begin{equation}\label{def:Ndiag}
m_N\,=\, m_N^\text{diag}\,=\,\text{diag}\,(m_{N_1},m_{N_2},m_{N_3})\,.
\end{equation}
The matrix $m_\nu$ can be diagonalised by the 
unitary matrix $U_{\text{MNS}}$, leading
to the following masses for the light physical states
\begin{equation}\label{physicalmasses}
U_\text{MNS}^T \,m_{\nu}\, U_\text{MNS} 
\,=\,m_{\nu}^\text{diag}
\,=\, \text{diag}\,(m_{\nu_1},m_{\nu_2},m_{\nu_3})\,.
\end{equation}
Here we will use the standard parameterisation for the $U_\mathrm{MNS}$,  
given by
\begin{equation}
U_\text{MNS}=
\left( 
\begin{array}{ccc} 
c_{12} \,c_{13} & s_{12} \,c_{13} & s_{13} \, e^{-i \delta} \\ 
-s_{12}\, c_{23}\,-\,c_{12}\,s_{23}\,s_{13}\,e^{i \delta} 
& c_{12} \,c_{23}\,-\,s_{12}\,s_{23}\,s_{13}\,e^{i \delta} 
& s_{23}\,c_{13} \\ 
s_{12}\, s_{23}\,-\,c_{12}\,c_{23}\,s_{13}\,e^{i \delta} 
& -c_{12}\, s_{23}\,-\,s_{12}\,c_{23}\,s_{13}\,e^{i \delta} 
& c_{23}\,c_{13}
\end{array} \right) \,.\, P\,,
\label{Umns}
\end{equation}
with
$P\,=\,\text{diag}\,(e^{-i\frac{\varphi_1}{2}},e^{-i\frac{\varphi_2}{2}},1)$,  
and where $c_{ij} \equiv \cos \theta_{ij}$,
$s_{ij} \equiv \sin \theta_{ij}$.
The parameters $\theta_{ij}$ are the neutrino flavour 
mixing angles, $\delta$ is the Dirac
phase and $\varphi_{1,2}$ are the Majorana phases. 

In view of the above, 
the seesaw equation, Eq.~(\ref{seesaw:light}),
can be solved for the neutrino Yukawa coupling $\lambda_\nu$
using the Casas-Ibarra parameterisation~\cite{Casas:2001sr} as 
\begin{equation}\label{seesaw:casas}
\lambda_\nu\,v_2\,=\, i \,\sqrt{m^\text{diag}_N}\, R \,
\sqrt{m^\text{diag}_\nu}\,  U^\dagger_{\text{MNS}}\,,
\end{equation}
where $R$ is a generic complex orthogonal $3 \times 3$ matrix that
encodes the possible extra neutrino mixings (associated with the
right-handed sector) in addition to those in the
$U_{\text{MNS}}$. $R$ can be parameterised 
in terms of three complex angles, $\theta_i$ $(i=1,2,3)$ as
\begin{equation}\label{Rcasas}
R\, =\, 
\left( 
\begin{array}{ccc} 
c_{2}\, c_{3} & -c_{1}\, s_{3}\,-\,s_1\, s_2\, c_3
& s_{1}\, s_3\,-\, c_1\, s_2\, c_3 \\ 
c_{2}\, s_{3} & c_{1}\, c_{3}\,-\,s_{1}\,s_{2}\,s_{3} 
& -s_{1}\,c_{3}\,-\,c_1\, s_2\, s_3 \\ 
s_{2}  & s_{1}\, c_{2} & c_{1}\,c_{2}
\end{array} 
\right)\,,
\end{equation}
with $c_i\equiv \cos \theta_i$, $s_i\equiv \sin\theta_i$. 
Eq.~(\ref{seesaw:casas}) is a convenient means of parameterising our ignorance 
of the full neutrino Yukawa couplings, while at the same time allowing
to accommodate the experimental data.
Notice that it is only valid at the 
right-handed neutrino scales $m_M$, so that the quantities appearing
in Eq.~(\ref{seesaw:casas}) are the renormalised ones, 
$m^\text{diag}_\nu\,(m_M)$ and $U_{\text{MNS}}\,(m_M)$.

In this study, we shall mainly focus on the simplest scenario, where
both heavy and light neutrinos are hierarchical, 
$m_{N_1}\, \ll\, m_{N_2}\, \ll \,m_{N_3}$ and  
$m_{\nu_1}\, \ll\, m_{\nu_2}\, \ll\, m_{\nu_3}$, and in particular, we
will assume a normal ordering of the light neutrinos.   
Thus, the masses $m_{\nu_{2,3}}$ 
can be written in terms of the lightest mass $m_{\nu_{1}}$ and   
the solar/atmospheric mass-squared differences as
$m_{\nu_2}^2\,=\, \Delta m_\text{sol}^2 \, + \, m_{\nu_1}^2$ and  
$m_{\nu_3}^2\,=\, \Delta m_\text{atm}^2 \, + \, m_{\nu_1}^2$.

In summary, when working in the $R$-matrix parameterisation, 
the 18 parameters of the seesaw mechanism are accounted by 
the three heavy neutrinos masses, $m_{N_i}$, the mass of
the lightest neutrino $m_{\nu_1}$ plus the two mass squared
differences $\Delta m_\text{sol}^2$ and $\Delta m_\text{atm}^2$, the
three mixing angles $\theta_{ij}$ and three CP violating phases
$\delta,\varphi_1,\varphi_2$ of the $U_{\text{MNS}}$ 
matrix, and the three complex angles $\theta_i$ of the matrix $R$. 
As mentioned in the Introduction, many of the latter parameters, namely those
associated with the heavy neutrino sector, are experimentally unreachable.
Nevertheless, it is possible to derive interesting bounds from the requirement
of successful thermal leptogenesis, and we proceed to do so in the following
sections.

\section{Flavour-dependent thermal leptogenesis with \\
  reheating}\label{Sec:FML} 
As recently pointed out~\cite{davidsonetal,nardietal,Abada:2006ea},
flavour can have a strong impact in baryogenesis via thermal
leptogenesis. 
The effects are manifest not only in the flavour-dependent CP
asymmetries, but also in 
the flavour-dependence of scattering processes in the thermal bath,
which can destroy a previously produced asymmetry.  
In fact, depending on the temperatures at which 
thermal leptogenesis takes place, and thus on 
which interactions mediated by the charged lepton Yukawa
couplings are in thermal equilibrium, 
flavour-dependent effects can 
have a strong impact on the estimation of the produced baryon 
asymmetry~\cite{barbieri}-\cite{Branco:2006ce}. For example, 
in the MSSM, for temperatures
between circa $(1+\tan^2 \beta)\times 10^{5} \: \mbox{GeV}$ and 
$(1+\tan^2 \beta)\times 10^{9} \: \mbox{GeV}$, the $\mu$
and $\tau$ Yukawa couplings are in thermal equilibrium and all
flavours in the Boltzmann equations are to be treated separately. For
instance, for $\tan \beta = 30$, this applies for temperatures below
about $10^{12}$ GeV (and above $10^{8} \: \mbox{GeV}$), a temperature range 
we will be subsequently considering. 
Moreover, in the full flavour-dependent treatment, lepton asymmetries are
generated in each individual lepton flavour. Processes which can wash
out these asymmetries are also flavour-dependent, i.e.\ the inverse
decays from electrons can only destroy the lepton asymmetry in the
electron flavour.  
We will address the latter issues in Section~\ref{bau:1}.

In thermal leptogenesis, the population of right-handed neutrinos
$N_1$ is produced from scattering processes in the thermal bath. To
generate the observed baryon asymmetry comparatively high
temperatures of the early Universe are required, and these should not
lie much below the mass of the lightest right-handed neutrino, 
$m_{N_1}$. Even under
optimal conditions, thermal leptogenesis demands 
$m_{N_1} \gtrsim 10^9$ GeV 
(for hierarchical light and heavy neutrinos)~\cite{Davidson:2002qv}.  
High temperatures compatible with thermal leptogenesis can arise in
the process of reheating which takes place 
after cosmic inflation~\cite{Guth:1980zm}.      
The temperature of the Universe at the end of reheating is referred to
as the reheat temperature $T_\mathrm{RH}$.  
However, particularly in locally supersymmetric theories,
$T_\mathrm{RH}$ is often constrained, as we will discuss in
Section~\ref{Sec:GravitinoProblem}.  

In the presence of such bounds on $T_\mathrm{RH}$, the requirement of
successful thermal leptogenesis imposes severe constraints on the
seesaw parameters. In our numerical calculations, and following 
Ref.~\cite{Giudice:2003jh}, we include 
reheating in the flavour-dependent Boltzmann 
equations~\cite{barbieri,davidsonetal,nardietal,Abada:2006ea}, 
generalised to the MSSM~\cite{Antusch:2006cw}, 
in a simplified but comparatively
model-independent way.
In particular, we assume that the lightest right-handed (s)neutrinos are
only produced by thermal scatterings during and after reheating.
Moreover, we neglect model-dependent issues such as the 
production of $N_1$ (and $\widetilde N_1$) 
during preheating, or from the decays of the scalar field responsible for
reheating.  
For completeness, the technical aspects of the Boltzmann equations 
are given in Appendix~\ref{App:BoltzmannReheating}. Further details on
the estimation of the produced baryon asymmetry using the 
flavour-dependent Boltzmann equations can be found
in Refs.~\cite{barbieri,davidsonetal,nardietal,Abada:2006ea,Antusch:2006cw}. 
Finally, and concerning the inclusion
of reheating, we refer the reader to Ref.~\cite{Giudice:2003jh}.

Let us now begin by reviewing (omitting technical aspects)
the procedure for estimating the baryon
asymmetry produced by thermal leptogenesis in the MSSM 
when reheating effects are included.

\subsection{Estimation of the produced baryon asymmetry}\label{bau:1}
The out-of-equilibrium decays of the heavy right-handed (s)neutrinos
give rise to flavour-dependent asymmetries in the (s)lepton sector,
which are then partly transformed via sphaleron conversion 
into a baryon asymmetry
$Y_B$\footnote{Throughout this study
$Y$ will always be used for quantities which are normalised to the
entropy density.}.
The final baryon asymmetry can be calculated
as~\cite{Antusch:2006cw}
\begin{eqnarray}\label{Eq:YB3f}
Y_B \,=\, \frac{10}{31}\, \sum_\alpha \hat Y_{\Delta_\alpha}\; ,
\end{eqnarray}
where $\hat Y_{\Delta_\alpha}\equiv Y_B/3 - Y_{L_\alpha}$ are the
total (particle and sparticle) $B/3 - L_\alpha$ asymmetries, with 
$Y_{L_\alpha}$ the lepton number densities in the flavour {\small $\alpha =
e, \mu,\tau$}.
The asymmetries $\hat Y_{\Delta_\alpha}$, which are conserved by sphalerons
and by the other MSSM interactions,
are then calculated by solving a set of coupled Boltzmann equations,
describing the evolution of the
number densities as a function of temperature.
We consider the simplest case of thermal leptogenesis, where only the
lightest right-handed neutrinos are produced in the thermal
bath\footnote{The limitations of this (and other)
approximation(s) will be discussed in Section~\ref{Sec:limitations}.}.
In the MSSM,
the asymmetries for the decay of the lightest right-handed
(s)neutrinos into (s)leptons of flavour {\small$\alpha$} (defined in
Eq.~(\ref{Eq:EpsMSSM_def})) satisfy
$\varepsilon_{1,\alpha} \!=\! \varepsilon_{\widetilde 1,\alpha}
\!=\!\varepsilon_{1, \widetilde \alpha} \!=
 \!\varepsilon_{\widetilde 1,\widetilde \alpha}$.
Thus, it is convenient to write the solutions of the Boltzmann
equations in terms of the
flavour-dependent decay asymmetries $\varepsilon_{1,\alpha}$
and flavour-dependent efficiency factors $\eta_{\alpha}$ as
\begin{eqnarray}\label{Eq:eta_aa}
\hat Y_{\Delta_\alpha}\, =\, \eta_{\alpha} \,
\varepsilon_{1,\alpha}\, \left[ \,
Y^{\mathrm{eq}}_{N_1}(T \gg m_{N_1}) + Y^{\mathrm{eq}}_{\widetilde
  N_1}(T \gg m_{N_1}) \,
\right]\,.\label{Eq:eta_aa_MSSM}
\end{eqnarray}
In the above,
$Y^{\mathrm{eq}}_{N_1}$ and  $Y^{\mathrm{eq}}_{\widetilde N_1}$ are the
number densities of the lightest right-handed neutrino and sneutrino in the
Boltzmann approximation (i.e.\ assuming common 
phase space densities for both fermions and scalars)
if they were in thermal equilibrium at $T \gg
m_{N_1}$,
\begin{eqnarray}\label{eq:Yeq}
 Y^{\mathrm{eq}}_{N_1}(T \gg m_{N_1}) \approx Y^{\mathrm{eq}}_{\widetilde N_1}
(T \gg m_{N_1}) \approx \frac{45}{ \pi^4
 g_* } \;,
 \end{eqnarray}
with $g_* = 228.75$ denoting the effective number of degrees of freedom.
While the equilibrium number densities mainly serve as a
normalisation, the relevant quantities are the decay asymmetries and
the efficiency factor, which we now proceed to specify.

\subsubsection{The flavour-dependent decay asymmetries 
$\boldsymbol{\varepsilon_{1,\alpha}}$}
In the basis where both the charged lepton and right-handed neutrino
mass matrices are diagonal, 
the asymmetries $\varepsilon_{1,\alpha}$ for the decay of the
lightest right-handed neutrino into lepton and Higgs doublets
(c.f.~Eq.~(\ref{Eq:EpsMSSM_def})) are given by~\cite{Covi:1996wh} 
\begin{eqnarray}\label{Eq:EpsMSSM}
\varepsilon_{1,\alpha} \,=\, 
\frac{1}{8\pi}\,
\frac{\sum_{J=2,3}\mathrm{Im}\left[
(\lambda_{\nu})_{1\alpha}\,
(\lambda_{\nu} \, \lambda^{\dagger}_{\nu})_{1J}\,
(\lambda_{\nu}^\dagger)_{\alpha J}
\right]}
{(\lambda_{\nu}^{\dagger} \, \lambda_{\nu})_{11}}\,
g\left(\frac{m_{N_J}^2}{m_{N_1}^2}\right) , 
\end{eqnarray}
where
\begin{eqnarray}
g(x) \,=\,\sqrt{x}\, \left[
\frac{2}{1-x} - \ln\left(\frac{1+x}{x}\right)\right]\,
\stackrel{x \gg 1}{\longrightarrow} \, - \frac{3}{\sqrt{x}}  \; .
\end{eqnarray}
For $m_{N_1} \ll m_{N_2},m_{N_3}$, and using the 
$R$-parameterisation (see Eq.~(\ref{seesaw:casas})), 
the decay asymmetries in the MSSM can be written as
\begin{eqnarray}\label{Eq:Epsa_R}
\varepsilon_{1,\alpha} \,=\, - \frac{3 \,m_{N_1}}{8\, \pi \,v_2^2} \,
\frac{\mathrm{Im} \left[\sum_{\beta \rho}\, 
m_{\nu_\beta}^{1/2}\, m_{\nu_\rho}^{3/2}\, (U_{\mathrm{MNS}})^*_{\alpha\beta} 
(U_{\mathrm{MNS}})_{\alpha\rho} \,R_{1 \beta } \,R_{1 \rho }
\right]}{\sum_\delta \,m_{\nu_\delta} \,|R_{1 \delta}|^2}\;.
\end{eqnarray}
Notice that there is no sum over {\small $\alpha$} in Eq.~(\ref{Eq:Epsa_R}),
which implies that both the $U_{\mathrm{MNS}}$ phases and a complex
$R$-matrix can contribute to the CP violation necessary for
leptogenesis. This has been recently pointed out by several 
authors~\cite{Abada:2006ea,Blanchet:2006be,Pascoli:2006ie,Branco:2006ce},
and is in direct contrast with the flavour-independent approximation,
where (working in the $R$-parameterisation) 
the $U_{\mathrm{MNS}}$ plays no role in the decay asymmetry 
$\varepsilon_1 = \sum_\alpha \varepsilon_{1,\alpha}$.

\subsubsection{The efficiency factors $\boldsymbol{\eta_{\alpha}}$}
\begin{figure}[t]
 \centering
 \includegraphics[scale=0.8,angle=0]{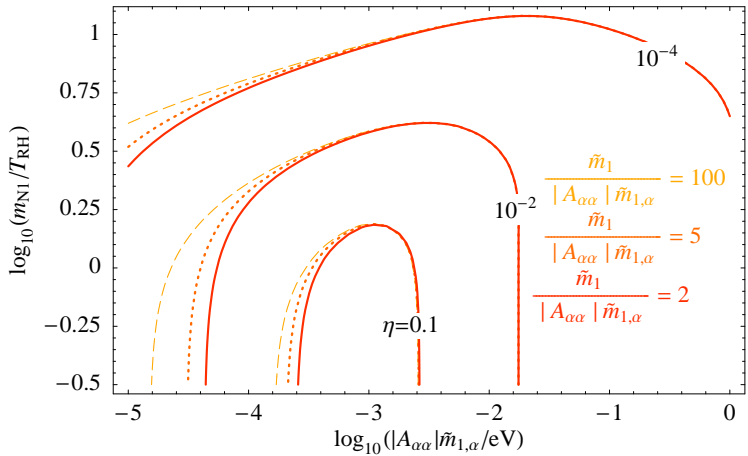}
 \caption{\label{fig:eta}
   Contour lines for the 
   flavour-dependent efficiency factor $\eta$ in the MSSM as a function of
   $|A_{\alpha\alpha}| \widetilde m_{1,\alpha}$ 
   (no sum over {\small$\alpha$}) and
   $m_{N_1}/T_\mathrm{RH}$, for fixed
   values of $\widetilde m_1/(|A_{\alpha\alpha}| \widetilde m_{1,\alpha})$.
   We take $ \widetilde m_1/(|A_{\alpha\alpha}| \widetilde m_{1,\alpha})= 2,5$
   and $100$, respectively
   corresponding to the solid, dotted and dashed contour lines. }
\end{figure}

As already mentioned,
the lepton asymmetries in each individual flavour are obtained
by solving the set of flavour-dependent Boltzmann equations, taking
into account reheating effects (c.f.~Appendix
\ref{App:BoltzmannReheating}).
Parameterising the solution of the Boltzmann equations as in
Eq.~(\ref{Eq:eta_aa}) implicitly defines the efficiency factor $\eta$.
In our approximation, $\eta$ is a function of the ratio
$m_{N_1}/T_{\mathrm{RH}}$, 
of the product $A_{\alpha\alpha} \,\widetilde m_{1,
\alpha}$ (no sum over {\small$\alpha$}), and of the total washout parameter
$\widetilde m_{1}$. The quantities
$\widetilde m_{1,\alpha}$ and $\widetilde m_{1}$ are
defined as
\begin{eqnarray}\label{Eq:mtildeaa}\label{eq:mtildeaa}
\widetilde{m}_{1,\alpha }\, \equiv\,
(\lambda_{\nu})_{1 \alpha}\,
(\lambda^{\dagger}_{\nu})_{\alpha 1}\,
\frac{v_{2}^2}{m_{N_1}}  \;  , \quad \quad
\widetilde{m}_1\, \equiv \,\sum_\alpha \widetilde{m}_{1,\alpha }\; .
\end{eqnarray}
If leptogenesis takes place at temperatures between about $(1+\tan^2
\beta)\times 10^{5} \: \mbox{GeV}$ and $(1+\tan^2 \beta)\times 10^{9} \:
\mbox{GeV}$, which is the
case we will consider in this study, $A$ is given as in
Ref.~\cite{Antusch:2006cw} (see Appendix~\ref{App:BoltzmannReheating}).
Here we will neglect the small off-diagonal elements of $A$, and use only the
leading diagonal entries, 
\begin{eqnarray}
A \approx \mbox{diag}(-93/110, -19/30,-19/30)\;.
\end{eqnarray}
In addition, there is also a slight dependence of $\eta$ on $\tan \beta$.
We also notice that 
the parameters $K$ and $K_\alpha$ (see Eq.~(\ref{Eq:Ka_mt}))
are often used instead of $\widetilde m_1$ and $\widetilde m_{1,\alpha}$.
Using the $R$-parameterisation, the washout parameters
$\widetilde m_{1,\alpha}$
can be written as
\begin{eqnarray}\label{Eq:mtildea_R}
\widetilde{m}_{1,\alpha } \,=\,
\big| \sum_{\beta} \,\sqrt{m_{\nu_\beta}} \,R_{1 \beta}
\,(U_\mathrm{MNS}^\dagger)_{\beta \alpha} \big|^2 \,.
\end{eqnarray}
The numerical results for the efficiency factor
as a function of $|A_{\alpha\alpha}| \widetilde m_{1,\alpha}$ and
$m_{N_1}/T_\mathrm{RH}$
are shown in Fig.~\ref{fig:eta}. As can be seen, the efficiency is indeed
optimal for values of $|A_{\alpha\alpha}| \widetilde m_{1,\alpha} \approx
m^*$, with $ m^*=
\sin^2 \beta \times 1.58 \times 10^{-3}$ eV~\cite{Antusch:2006cw} (see
Eqs.~(\ref{Eq:Kaa}, \ref{Eq:Ka_mt})), and quickly drops for
either smaller or larger $|A_{\alpha\alpha}| \widetilde m_{1,\alpha}$.
From Fig.~\ref{fig:eta}, it is apparent that
for $|A_{\alpha\alpha}| \widetilde m_{1,\alpha} < m^*$, 
the efficiency exhibits a less pronounced drop if
$\widetilde m_{1} /(|A_{\alpha\alpha}| \widetilde m_{1,\alpha}) \gg 1$.
Moreover, one verifies that the
efficiency is strongly reduced if $m_{N_1} \gg T_\mathrm{RH}$, for
instance by more than three orders of magnitude if $m_{N_1} \approx
10 \times T_\mathrm{RH}$. This stems from
our assumption that $N_1$ (and $\widetilde N_1$) are 
exclusively produced 
from thermal scatterings during and after reheating.
With respect to Fig.~\ref{fig:eta},
let us finally notice that even though the curves were obtained for an
example of $\tan \beta=30$, the results are
approximately the same for other moderate (and even large)
values of $\tan \beta$. Likewise, the contour lines for larger $\widetilde
m_1/(|A_{\alpha\alpha}| \widetilde m_{1,\alpha})$ 
look virtually the same as for
$\widetilde m_1/(|A_{\alpha\alpha}| \widetilde m_{1,\alpha}) = 100$.

One important difference between the flavour-dependent treatment
and the flavour-independent approximation is that in the former case
the total baryon asymmetry is the sum of the distinct
individual lepton asymmetries,
weighted by the corresponding efficiency factor, as in
Eq.~(\ref{Eq:eta_aa}).
Therefore, the total baryon number is in general not proportional to
the sum over the individual
CP asymmetries $\varepsilon_1 = \sum_\alpha \varepsilon_{1,\alpha}$,
as assumed in the flavour-independent approximation, where
flavour is not considered in the Boltzmann equations.
Moreover, the flavour-dependent efficiency factors $\eta_\alpha$ are in
general not equal to each other, and can strongly differ from the
common efficiency factor $\eta$
(itself a function of the common washout parameter
$\widetilde m_{1}$ of the
flavour-independent approximation).
Taking all the latter into account can lead to dramatic
differences regarding the estimates of the produced baryon asymmetry
in models of neutrino masses an mixings~\cite{Abada:2006ea,Antusch:2006cw}.

\subsubsection{Limitations and approximations of the 
computation}\label{Sec:limitations}
As this point, it is pertinent to summarise the approximations and
assumptions leading to the calculation of the produced baryon
asymmetry.  
 
For the right-handed neutrino sector, we assume a hierarchical
mass spectrum, which means that we do not take into account the
possibility of resonant effects in
leptogenesis~\cite{res}. Furthermore, we assume that 
only decays of $N_1$ 
(and $\widetilde N_1$) significantly contribute to the final
asymmetry, 
thus neglecting contributions from the decays of $N_2,N_3$ (and
$\widetilde N_2,\widetilde N_3$). This is justified
when either the washout processes mediated by the lighter right-handed
(s)neutrinos are 
sufficiently active for each flavour, and thus indeed destroy the
asymmetries arising from the decays of the heavier right-handed (s)neutrinos,
or when $T_\mathrm{RH} \ll m_{N_2},m_{N_3}$.
It is important to notice that under certain conditions
the heavier right-handed neutrinos can also
play a role for leptogenesis~\cite{Vives:2005ra},  
and should in principle be included. 

We include reheating in the flavour-dependent MSSM Boltzmann
equations in a simplified, but comparatively model-independent way, 
following Ref.~\cite{Giudice:2003jh}.
We notice that in specific models of reheating after inflation,
the prospects for leptogenesis can be 
different\footnote{Constraints on the seesaw parameters differ significantly
  if, for instance, the inflaton predominantly decays into right-handed
  (s)neutrinos~\cite{nonthermalLG}, or if one 
  works in scenarios where the inflaton is one
  of the right-handed (singlet) sneutrinos~\cite{SneutrinoInflation}.}. 
In particular, we assume 
that the lightest right-handed (s)neutrinos
are only produced by thermal scatterings during and after reheating.
We also neglect the possibility of producing $N_1$ (and
$\widetilde N_1$) from the decays of the scalar field $\phi$ responsible for
reheating, or of producing $N_1$ (and $\widetilde N_1$) during
preheating.
It is further assumed that the maximally 
reached temperature during the reheating
process, $T^\mathrm{max}$, as well as the mass of $\phi$, are much larger
than $m_{N_1}$.
Furthermore, we neglect the potential implications of supersymmetric flat
directions for reheating and thermal leptogenesis, which are still
under controversial discussion~\cite{flatdir}.

When solving the Boltzmann equations for flavour-dependent
leptogenesis, we focus our attention on the case where,
during the relevant temperatures for leptogenesis,
the interactions mediated by each of the charged lepton Yukawa
couplings are either fully in equilibrium, or out of equilibrium. 
In the MSSM, for values of $m_{N_1}$ around 
$(1 + \tan^2 \beta) \times 10^9$ GeV,  
the reactions induced by the muon Yukawa coupling are close to 
equilibrium and the quantum oscillations of the asymmetries may not
have been dumped fast enough to be neglected. To take these
effects into account, the Boltzmann 
equations may be generalised so to include quantum 
oscillations~\cite{davidsonetal}.  
In this study, we chose $\tan \beta$ sufficiently 
large so that we can safely assume that the charged $\mu$ and $\tau$
Yukawa couplings are in thermal equilibrium, and that
all flavours in the Boltzmann
equations can be treated separately.  
Furthermore, we neglect the small off-diagonal elements of the matrix $A$, 
which appears in the washout terms of the Boltzmann equations 
for $\hat Y_{\Delta_\alpha}$ (c.f.~Appendix \ref{App:BoltzmannReheating}). 
We note that this approximation is crucial if one wants 
to introduce an efficiency factor $\eta$, which is 
a function of the ratio $m_{N_1}/T_{\mathrm{RH}}$, 
of the product $A_{\alpha\alpha} \,\widetilde m_{1,
\alpha}$ (no sum over {\small$\alpha$}), and of the total washout parameter
$\widetilde m_{1}$.
We have numerically verified that in the regions of interest
for this study, including the off-diagonal elements has only a small effect
of increasing the produced BAU by about 20\%.   

Following Ref.~\cite{pedestrians}, 
in our numerical computations we only include processes in the
thermal bath mediated by neutrino and top Yukawa couplings. 
This means that we neglect $\Delta
L = 1$ scatterings involving gauge 
bosons~\cite{Pilaftsis:2003gt,Giudice:2003jh}, thermal 
corrections~\cite{Giudice:2003jh} and possible effects from 
spectator processes~\cite{Buchmuller:2001sr}, but that we do take
into account corrections from
renormalisation group (RG) running between the electroweak scale 
and $m_{N_1}$~\cite{barbieri,Antusch:2003kp} (for which we 
use the software package REAP~\cite{Antusch:2005gp}).  
Regarding the pole mass of the top quark, we take the value 
$m_t^\text{pole}=174.2$~GeV~\cite{PDG}. We stress that
the uncertainties in this value can have a strong influence on the RG 
evolution of the relevant parameters (namely the neutrino masses). 
Thus, the latter can  
provide a significant
source of uncertainties in the BAU estimates. 
The renormalised value of the top Yukawa coupling (at energies
$m_{N_1}$), also directly affects the strength of the $\Delta L = 1$
scatterings. Furthermore, let us notice that we also 
neglect $\Delta L=2$ scatterings, which is a good
approximation as long as 
$|A_{\alpha \alpha}| \widetilde m_{1,\alpha}/m^* \gg 10 \times
(m_{N_1} / 10^{14} \:\mbox{GeV})$~\cite{Abada:2006ea}. 

Finally, our
estimates of the produced BAU are based on a set of coupled Boltzmann
equations, which only provide a classical approximation to the
Kadanoff-Baym equations. Quantum effects for thermalization have been
ignored in our analysis~\cite{Lindner:2005kv}. 
Other approximations have led to the present simplified 
form of the Boltzmann equations:
for instance, it is usually assumed that elastic scattering rates are fast and
that the phase space densities for both fermions and scalars can be
approximated as
$f(E_i,T)=(n_{i}/n^{\mathrm{eq}}_{i}) e^{-E_i/T}$, where
$n_i^\mathrm{eq} = \tfrac{g_i}{2\pi} T m_i^2 K_2 (m_i/T)$, 
with $g_i$ being the number of degrees of freedom. 
Accordingly, we also use the equilibrium number densities in this
so-called Boltzmann approximation.

All the above mentioned approximations (as well as others we have not
explicitly discussed) 
will ultimately translate in potentially relevant
theoretical uncertainties when estimating the value of the BAU. Thus,
it is important to bare in mind that one may be either over- or
under-estimating $n_\text{B}/n_\gamma$, so that caution should be
exerted when deciding upon the BAU viability of a given SUSY seesaw scenario. 

\subsection{Constraints on the reheat temperature and the 
gravitino problem}\label{Sec:GravitinoProblem}

The predictions for the baryon asymmetry arising from a given seesaw
scenario can be severely compromised due to constraints on
$T_\mathrm{RH}$.
One important example of such constraints emerges in 
locally supersymmetric theories, and stems from the 
decays of thermally produced gravitinos~\cite{gravitinoproblem,Kohri:2005wn}.  
In this class of SUSY models, and assuming 
a low-energy MSSM with R-parity conservation, either the gravitino is 
the lightest supersymmetric particle (LSP) and is thus stable, or else
it will ultimately decay into the LSP. Two generic problems arise from
these decays, and are the following.

Firstly, gravitinos can decay late, after the Big Bang nucleosynthesis (BBN) 
epoch, and potentially spoil the success of
BBN~\cite{gravitinoproblem,thermal}. This leads to upper bounds on the
reheat temperature which depend on  
the specific supersymmetric model as well as on the mass of the 
gravitino~\cite{Kohri:2005wn}. For gravitino masses in the TeV region,
the gravitino BBN problem practically  
precludes the possibility of thermal leptogenesis.
However, with a heavy gravitino (roughly above $100$ TeV), 
the BBN problems can be nearly avoided.  
If one considers the gravitino mass as a free parameter, 
one can safely avoid the latter constraints for any given reheat temperature. 

Secondly, the decay of a gravitino produces one LSP, which has an impact 
on the relic density of the latter. 
The number of thermally produced gravitinos increases with the reheat 
temperature, and we can estimate the contribution to the dark matter (DM) 
relic density arising from
non-thermally produced LSPs via gravitino decay (for heavy gravitinos)
as~\cite{thermal,Kohri:2005wn} 
\begin{equation}
\Omega^{\mathrm{non-th}}_\mathrm{LSP} \,h^2 \approx 0.054 
\left( \frac{m_\mathrm{LSP}}{100 \, \text{GeV}}\right) 
\left( \frac{T_\mathrm{RH}}{10^{10}\,\text{GeV}} \right) ,
\end{equation}
which depends on the LSP mass, $m_\mathrm{LSP}$, as well as on 
$T_\mathrm{RH}$.     
Taking the relic density 
bound $\Omega^{\mathrm{non-th}}_\mathrm{LSP} h^2 \le
\Omega_\mathrm{DM} h^2 \lesssim 0.13$ from the Wilkinson Microwave
Anisotropy Probe (WMAP)~\cite{Spergel:2006hy}, 
we are led to an upper bound on the reheat temperature of
\begin{equation}\label{TRHbound}
T_{\mathrm{RH}}\, \lesssim \,2.4 \times 10^{10} \, \text{GeV} \,
\left(  \frac{100 \: \text{GeV}}{m_\mathrm{LSP} }\right)  .
\end{equation}
For masses of the LSP (assuming this to be the
lightest neutralino) in the range 100 GeV$\, - \, 150$ GeV we obtain
an estimated upper bound on the reheat temperature of approximately 
$T_{\mathrm{RH}} \lesssim 2 \,\times\, 10^{10} \,
\text{GeV}$. Naturally, heavier LSP masses lead to more severe bounds
on $T_\mathrm{RH}$.

There are other frameworks where, although still viable, 
thermal leptogenesis is 
significantly constrained by bounds on $T_\mathrm{RH}$. This can occur for
scenarios with stable gravitinos, i.e.\ where gravitinos are the
LSP. In many cases the bounds on the reheat temperature strongly 
depend on the
model under consideration, for instance on the properties of the
next-to-lightest supersymmetric particle (NLSP). 
For example, recent studies of models with gravitino
LSP~\cite{Roszkowski:2004jd} have found the following 
bounds for the Constrained Minimal Supersymmetric Standard Model (CMSSM),
\begin{eqnarray}\label{TRHbound2}
T_{\mathrm{RH}} \lesssim 4 \,\times\, 10^{9} \, \text{GeV}\,,
\end{eqnarray}
while a scenario with gaugino-mediated supersymmetry breaking and
sneutrino NLSP (stau NLSP) implies~\cite{hep-ph/0609142}
\begin{eqnarray}\label{TRHbound3}
T_{\mathrm{RH}} \lesssim 7 \,\times\, 10^{9} \, \text{GeV}\;\;
(T_{\mathrm{RH}} \lesssim 3\,\times\, 10^{9} \, \text{GeV})\; . 
\end{eqnarray} 

In the subsequent numerical analysis, 
and as examples, we will take into account the following
bounds, $T_{\mathrm{RH}} \le 2
\,\times\, 10^{10} \, \text{GeV}$ and $T_{\mathrm{RH}} \le 5
\,\times\, 10^{9} \, \text{GeV}$, respectively representative of a
mildly and a strongly constrained case for thermal
leptogenesis.

\section{Constraints on the seesaw parameters}\label{sec:constraints}

After having gone through the most relevant aspects of the computation
of the BAU, let us now proceed to discuss the constraints
on the several seesaw parameters which arise from assuming
that the baryon asymmetry is generated by flavour-dependent thermal
leptogenesis.  

The parameters of the $U_\text{MNS}$ matrix, as well as the 
two mass squared differences, are presently 
constrained by neutrino oscillation experiments. 
From them we know that $s_{23} \approx c_{23} \approx \sqrt{1/2}$,
$s_{12} \approx \sqrt{1/3}$ (and $c_{12} \approx \sqrt{2/3}$), while
$\theta_{13}$ is only bounded from above, $\theta_{13} \lesssim
11.5^\circ$ (at $3 \sigma$ confidence level)~\cite{Maltoni:2004ei}. 
Regarding the Dirac and Majorana phases
($\delta$, $\varphi_1$ and $\varphi_2$),
no experimental data is yet available.
For the case of hierarchical neutrinos, we have that 
$m_{\nu_2} \approx \sqrt{\Delta m^2_\text{sol}} \approx0.01$ eV 
and $m_{\nu_3}\approx \sqrt{\Delta m^2_\text{atm}} \approx 0.05$ eV.
On the other hand, parameters like the heavy neutrino masses 
$m_{N_i}$, and the
mixing angles involving the heavy Majorana neutrinos (i.e. the complex
$R$-matrix angles $\theta_i$), are experimentally unreachable.

The main focus of this work is to address the constraints on mixing
angles and CP violating phases of both the 
light and heavy neutrino sectors. However,
and especially when reheating effects are taken into account,
interesting bounds for the heavy and light
neutrino masses can also be derived. 
We begin our discussion by briefly re-analysing the latter constraints for 
the case of the MSSM, considering flavour-dependent effects.

\subsection{Heavy and light neutrino masses}

Let us start with general considerations regarding
bounds on the light and heavy neutrino masses from thermal
leptogenesis in the MSSM, when flavour effects are taken into account. 
From Eqs.~(\ref{Eq:Epsa_R}, \ref{Eq:mtildea_R}), it is clear that 
within our framework (and approximations), the masses $m_{N_2}$ and
$m_{N_3}$ are not constrained by thermal leptogenesis. 

In flavour-dependent leptogenesis, the decay asymmetries are
constrained by\footnote{For simplicity,
  we present the discussion for normal mass
  ordering. For inverse ordering, $m_{\nu_3}$ is
  replaced by $m_{\nu_2}$.}~\cite{Abada:2006ea} 
\begin{eqnarray}\label{Eq:BoundOnEpsa}
\varepsilon_{1,\alpha} \,\le \,\frac{3\,m_{N_1}}{8 \pi\, v_2^2} \,  
m_{\nu_3}\, \left(\frac{\widetilde m_{1,\alpha}}{\widetilde m_1}\right)^{1/2} 
\,\equiv \,\varepsilon^{\mathrm{max}}_1 \, 
\left(\frac{\widetilde m_{1,\alpha}}{\widetilde m_1}\right)^{1/2} .
\end{eqnarray}
In the above equation, and for hierarchical light neutrino masses, 
$\varepsilon^{\mathrm{max}}_1 = 3\,m_{N_1} \,m_{\nu_3}/(8 \pi\,
v_2^2)$ is the maximally possible value, both in the flavour-independent
approximation and in the flavour-dependent treatment. 
Regarding washout, in the type-I seesaw, the
flavour-independent washout parameter satisfies
\cite{Buchmuller:2003gz}  
\begin{eqnarray}\label{Eq:mtilde_and_m1}
\widetilde m_1 \,\ge\, m_{\nu_1} \; ,
\end{eqnarray} 
whereas the flavour-dependent washout parameters $\widetilde
m_{1,\alpha}$ are generically not constrained.  
In the flavour-independent approximation,
Eq.~(\ref{Eq:mtilde_and_m1}) leads to a dramatically more restrictive
bound on $\varepsilon_1 = \sum_\alpha 
\varepsilon_{1,\alpha}$~\cite{Davidson:2002qv}, and finally
even to a bound on the neutrino mass scale~\cite{Buchmuller:2003gz}. 
We also note that for quasi-degenerate light neutrino masses, an
optimal washout parameter\footnote{From here on, and regarding
  analytical discussions, we will assume
  $|A_{\alpha \alpha}| \widetilde m_{1,\alpha}\approx \widetilde
  m_{1,\alpha}$, thus neglecting the $\mathcal{O}(1)$
  quantity $A_{\alpha \alpha}$. The latter is included in the
  numerical computations.}
$\widetilde m_{1,\alpha}\approx m^*$ is
possible, but it however implies that the decay asymmetries are
suppressed by at least a factor 
$\sqrt{m^*/m_{\nu_1}}$ when compared to the optimal
value $\varepsilon^{\mathrm{max}}_1$
(c.f.~Eq.~(\ref{Eq:BoundOnEpsa})).\footnote{For
  instance, in the type-II seesaw, where an additional direct mass
  term for the light
  neutrinos from SU(2)$_\mathrm{L}$-triplets is present, this
  suppression can be avoided and the maximal decay asymmetry
  $\frac{3\,m_{N_1}}{8 \pi} \, m_{\nu_3}$ (for normal mass ordering)
  can be realised for quasi-degenerate neutrinos
  \cite{Antusch:2004xy}. It is easy to see that this holds true also
  in the flavour-dependent case. For instance, in ``type-II upgraded''
  seesaw models (see e.g.~\cite{Antusch:2005tu}), this bound can be nearly
  saturated easily.}   
Concerning the decay asymmetries, we will see that the suppression
factor $\sqrt{\widetilde m_{1,\alpha}/\widetilde m_1}$ 
has further interesting implications also for the case of hierarchical
neutrino masses.  

Using the upper bound on the decay asymmetry of  
Eq.~(\ref{Eq:BoundOnEpsa}) for the case of 
hierarchical light neutrinos, and
assuming an optimal efficiency, it is possible to estimate the baryon
asymmetry. Even without reheating, the comparison of the estimated
value with the observed BAU by WMAP~\cite{Spergel:2006hy}, 
\begin{equation}
n_\mathrm{B} /n_\gamma
 \,\approx\, (6.10\,\pm\,0.21)\,\times\,10^{-10}\,, 
\quad  \text{or}\quad 
Y_B\,\approx\, (0.87\,\pm\,0.03)\,\times\,10^{-10}\,,
\end{equation}
allows to obtain  a lower bound on
the mass $m_{N_1}$ of the lightest right-handed
neutrino~\cite{Davidson:2002qv}. 

\begin{figure}[t]
 \centering
 \includegraphics[scale=0.8,angle=0]{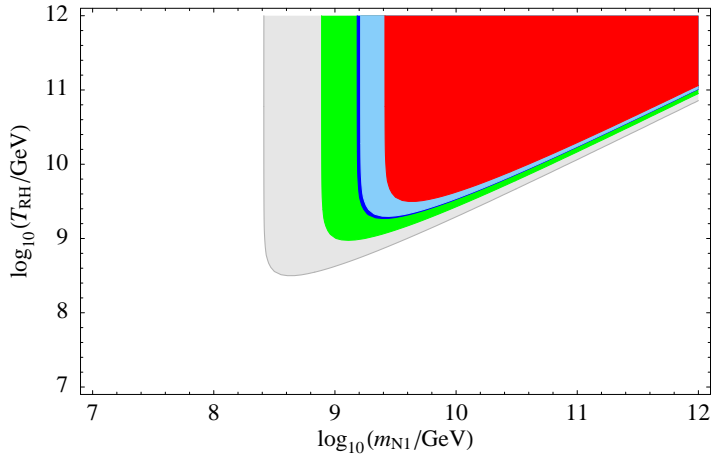}
 \caption{\label{fig:Bound_mN1_TRH}
Isosurfaces corresponding to the maximally possible BAU as function
of $m_{N_1}$ and $T_{\mathrm{RH}}$. 
From small to large $m_{N_1}-T_{\mathrm{RH}}$ regions, 
the associated BAU ranges are: 
$n_\text{B}/n_\gamma\in [10^{-10}, \,3 \times 10^{-10}]$,
$n_\text{B}/n_\gamma
\in [3 \times 10^{-10}, \,5.9 \times 10^{-10}]$, $n_\text{B}/n_\gamma
\in [5.9 \times 10^{-10}, \,6.3 \times 10^{-10}]$, $n_\text{B}/n_\gamma
\in [6.3 \times 10^{-10}, \,10^{-9}]$ and 
$n_\text{B}/n_\gamma \gtrsim 10^{-9}$. The corresponding colour code
is grey, green, dark blue (WMAP), light blue and red, respectively. 
}
\end{figure}

Clearly, the maximal BAU that can be generated depends on
both $m_{N_1}$ and $T_{\mathrm{RH}}$. The combined constraints on these
quantities are shown in Fig.~\ref{fig:Bound_mN1_TRH}. Leading to the latter,
we have considered a normal hierarchical spectrum of light neutrinos.
We have also assumed a maximal decay asymmetry as in
Eq.~(\ref{Eq:BoundOnEpsa}) and 
an optimal efficiency for a given
$m_{N_1}/T_{\mathrm{RH}}$. 
From Fig.~\ref{fig:Bound_mN1_TRH}, let us finally point out that in order to  
obtain BAU compatible with the WMAP range (represented in dark blue),
the minimal values for the reheat temperature and for $m_{N_1}$ are
$T^\mathrm{min}_{\mathrm{RH}} \approx 1.9\times 10^9$ GeV and
$m^\mathrm{min}_{N_1} \approx 1.5 \times 10^9$ GeV. 
Moreover, in the presence of an upper bound on the reheat temperature, 
there is also an upper bound on $m_{N_1}$, stemming from the
dramatic loss of efficiency occurring when $m_{N_1} \gg
T_{\mathrm{RH}}$, as it was shown in Fig.~\ref{fig:eta}. 
For instance, $T_\text{RH} \lesssim 2
\times 10^{10}$ GeV imposes 
$m^\text{max}_{N_1} \approx 1.4 \times 10^{11}$ GeV, while
$T_\text{RH} \lesssim 5 \times 10^{9}$ GeV yields 
$m^\text{max}_{N_1} \approx 1.9 \times 10^{10}$
GeV, leading to viability windows for the mass of the lightest right-handed
neutrino.

\begin{figure}[t]
 \centering
 \includegraphics[scale=0.8,angle=0]{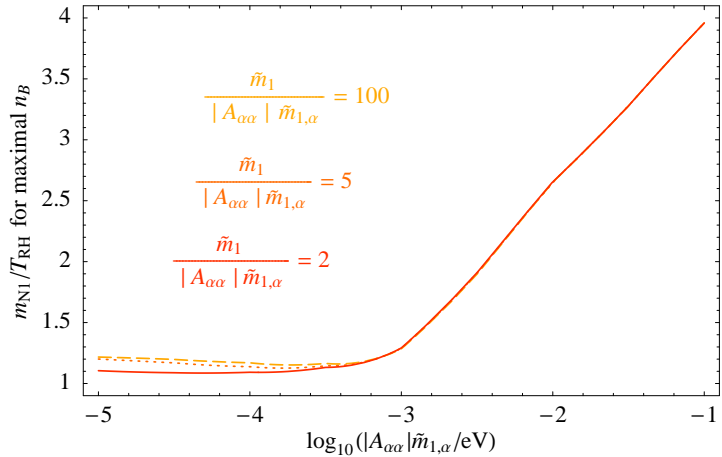}
 \caption{\label{fig:mN1opt}
Value of the ratio $m_{N_1}/T_\mathrm{RH}$ which yields the maximally
possible baryon asymmetry, as a function of $|A_{\alpha \alpha}| \widetilde
m_{1,\alpha}$. Solid, dotted, and dashed lines correspond to
$\widetilde m_{1}/(|A_{\alpha\alpha}|\,\widetilde m_{1,\alpha})=2, 5$ 
and $100$, respectively.      
}
\end{figure}

In the following analysis, whenever we present numerical examples
regarding constraints on the seesaw parameters, we will vary
$m_{N_1}$, and  select the value for which the produced baryon
asymmetry is maximal. Typically, this corresponds to $m_{N_1}$ around
(or slightly above) $T_{\mathrm{RH}}$, as illustrated in
Fig.~\ref{fig:mN1opt}. 
In order to reduce the BAU associated with a given choice of
parameters, one can simply vary $m_{N_1}$. Lowering $m_{N_1}$ such
that $m_{N_1} \ll m^*$ leads to a regime where $n_\text{B}/n_\gamma$
decreases with decreasing $m_{N_1}$. On the other hand, it is also
possible to increase $m_{N_1}$, taking values $m_{N_1} \gg
T_\text{RH}$, since then the strong washout leads to an important
reduction in the BAU.

\begin{figure}[t]
 \centering
 \includegraphics[scale=0.8,angle=0]{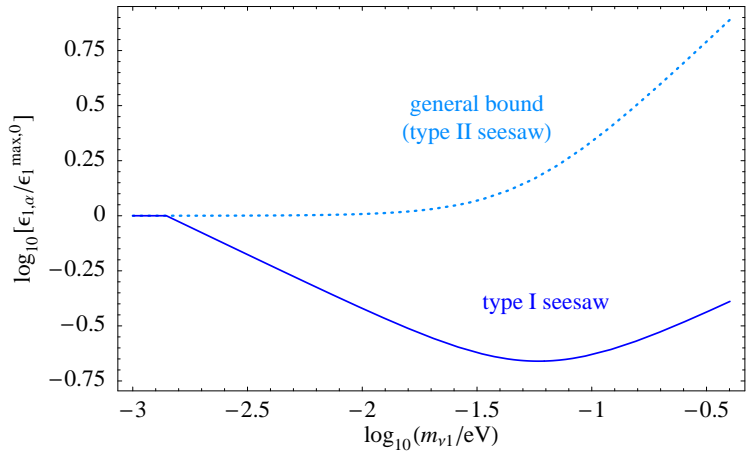}
 \caption{\label{fig:epsbound}
Bound on the decay asymmetry $\varepsilon_{1,\alpha}$ 
as a function of $m_{\nu_1}$. The  washout
parameter $|A_{\alpha \alpha}|
\widetilde m_{1,\alpha}$ is fixed to $m^*=\sin^2 \beta \times
1.58 \times 10^{-3}$ (close to optimal), and the asymmetry is 
normalised to its maximal value
$\varepsilon^{\mathrm{max,}0}$, obtained for a (normal)
hierarchical mass spectrum of light neutrinos with $m_{\nu_1}=0$. 
The bound of the decay asymmetry in the type-I seesaw (solid line) 
decreases with increasing neutrino mass scale and remains 
below the bound for hierarchical masses for the depicted 
values of $m_{\nu_1}<0.4$ eV. 
For comparison, the general bound, which can be saturated in the
type-II seesaw scenario, is also displayed (dotted line). 
The latter increases with increasing neutrino mass scale. 
We have used $\tan
\beta = 30$ and $m_{N_1}=10^{10}$ GeV. 
}
\end{figure}

Finally, and before concluding this discussion, let us comment on the
bounds for the masses of the light neutrinos. At present, 
the absolute neutrino mass scale $m_{\nu_1}$ (for normal mass
ordering) is only experimentally constrained by Tritium
$\beta$-decay, $0\nu\beta\beta$-decay and cosmology, to be 
roughly below $0.5$ eV~\cite{Aalseth:2004hb}. 
In general, flavour-dependent leptogenesis will not
provide any additional constraints on $m_{\nu_1}$.
As mentioned before, in the present study we focus on hierarchical
light neutrinos. 
In this case, it is nevertheless interesting to point
out that with a bound $T_{\mathrm{RH}} \le 5 \times 10^{9}$ GeV or
$T_{\mathrm{RH}} \le 2 \times 10^{10}$ GeV, we have 
numerically verified that increasing the neutrino mass scale towards 
a quasi-degenerate light neutrino mass spectrum leads
to a reduction of the BAU-allowed regions of the parameter space.
This is essentially due to two reasons. 
Firstly, although there is no bound on $\widetilde m_{1,\alpha}$, its
typical values are of the order of
$m_{\nu_1},m_{\nu_2}$ and $m_{\nu_3}$. Therefore, only for strongly
hierarchical light neutrino masses can $m_{\nu_1}$ set the right
scale for an optimal washout parameter. 
Secondly, slightly increasing $m_{\nu_1}$ towards a
quasi-degenerate spectrum does not significantly enhance
$\varepsilon_1^\mathrm{max}$, but does reduce each of 
the decay asymmetries due to
a factor of $\sqrt{\widetilde m_{1,\alpha}/\widetilde 
m_1} \approx \sqrt{m^*/\widetilde m_1} <
\sqrt{m^*/ m_{\nu_1}}$, enforcing optimal washout $\widetilde m_{1,\alpha}
\approx m^*$ (using Eq.~(\ref{Eq:mtilde_and_m1})).  
This is illustrated in Fig.~\ref{fig:epsbound}, 
where we display the bound on the decay asymmetry 
$\varepsilon_{1,\alpha}$, normalised to the maximal decay asymmetry
 $\varepsilon^{\mathrm{max,}0}=\frac{3\,m_{N_1}}{8 \pi\, v_2^2} \,
\sqrt{\Delta m^2_\text{atm}}\,$ for a (normal) hierarchical mass spectrum of
light neutrinos. The washout parameter $\widetilde m_{1,\alpha}$ has
been fixed to $m^*=\sin^2 \beta \times 1.58 \times 10^{-3}$ (close to
its optimal value).  

\subsection{Mixing angles and CP phases}
Here we will discuss how the constraints arising from the reheat
temperature can affect the washout and efficiency factors, and in turn
favour/disfavour choices for the mixing angles and CP violating phases. 
We also analyse the
impact of the latter constraints regarding the flavour-dependent CP
asymmetries, and investigate some illustrative limits regarding
the $R$-matrix angles, $\theta_i$.\footnote{We would like to note that our model-independent results, presented in terms of the $R$-matrix parameterisation, can be understood as constraints on whole equivalence classes of neutrino mass models (see e.g.\ \cite{King:2006hn}).}

\subsubsection{Favoured parameter regions with optimal washout}
\label{Sec:Washout}       

With strong constraints on the reheat temperature like
$T_{\mathrm{RH}} \le 5 \times 10^{9}$ GeV or $T_{\mathrm{RH}} \le 2
\times 10^{10}$ GeV (motivated by
Eqs.~(\ref{TRHbound}-\ref{TRHbound3})),  
and a lower bound on $m_{N_1}$ of about $10^9$
GeV, only a rather small window of $m_{N_1}$ values remains allowed 
(c.f.\ Fig.~\ref{fig:Bound_mN1_TRH}).  
This means that in order to produce enough baryon asymmetry, at least
one of the efficiency factors $\eta_\alpha = \eta (A_{\alpha \alpha} \widetilde
m_{1,\alpha}, \widetilde m_1, T_{\mathrm{RH}}/m_{N_1})$ should be
close to optimal (i.e., $\widetilde m_{1,\alpha}$ must not differ
much from $m^*$) and, in addition, the corresponding decay asymmetries
should approach $\varepsilon_1^\mathrm{max}$.    

An optimal efficiency $\eta$ can be achieved for $\widetilde m_{1,\alpha}
\approx m^* = \sin^2 \beta \times 1.58 \times 10^{-3}$ eV. 
Note that $m^*$ is much
smaller than $m_{\nu_3}$ and, even though $m_{\nu_2} < m_{\nu_3}$, $m^*$
is still significantly smaller than $m_{\nu_2}$. For hierarchical
light neutrino masses, apart from having $m_{\nu_1} \ll m_{\nu_2} <
m_{\nu_3}$, $m_{\nu_1}$ 
is unconstrained and can vary in a large range.  
For instance, it is possible  to choose $m_{\nu_1}$ close to $m^*$, 
such that $\widetilde m_{1,\alpha} \approx m_{\nu_1}$ 
would ensure an optimal value for the washout parameter.  
In the following numerical analysis, we will set $m_{\nu_1} =
10^{-3}\:\mbox{eV} \approx m^*$ as an illustrative example. 

To begin the analysis, let us start by writing the washout parameters
explicitly in terms of the seesaw parameters, using the $R$-matrix
parameterisation. One has 
\begin{eqnarray}\label{Eq:mt_1}
\widetilde m_{1,e} &=&
\big| 
e^{\frac{i}{2}\varphi_1}\, c_2\,c_3 \, c_{12} \,c_{13}\,\sqrt{m_{\nu_1}} 
+ 
s_3 (e^{i \delta} \,s_1 \,s_{13}\,\sqrt{m_{\nu_3}} - 
e^{\frac{i}{2}\varphi_2} \,c_1 \,s_{12} \,c_{13} \,\sqrt{m_{\nu_2}})
\nonumber \\ 
&& +
s_2 c_3 (
- e^{i \delta} \,c_1 \,s_{13}\,\sqrt{m_{\nu_3}}
- e^{\frac{i}{2}\varphi_2} \, s_1\,s_{12} \,c_{13} \,\sqrt{m_{\nu_2}} 
)
\big|^2 , \\\label{Eq:mt_2}
\widetilde m_{1,\mu} &=&
\big| 
- e^{\frac{i}{2}\varphi_1}\, c_2\,c_3 \, s_{12}\,c_{23} \,\sqrt{m_{\nu_1}} 
+ s_3 (s_1 \,s_{23}\,\sqrt{m_{\nu_3}} 
- e^{\frac{i}{2}\varphi_2} \,c_1 \,c_{12} \,c_{23} \,\sqrt{m_{\nu_2}})
\nonumber \\ 
&& +
s_2 c_3 (
- c_1 \,s_{23}\,\sqrt{m_{\nu_3}}
- e^{\frac{i}{2}\varphi_2} \, s_1\,c_{12} \,c_{23} \,\sqrt{m_{\nu_2}} 
) 
+ {\cal O}(\theta_{13})
\big|^2 , \\\label{Eq:mt_3}
\widetilde m_{1,\tau} &=&
\big| 
e^{\frac{i}{2}\varphi_1}\, c_2\,c_3\, s_{12} \, s_{23} \,\sqrt{m_{\nu_1}} 
+ s_3 (s_1 \,c_{23}\,\sqrt{m_{\nu_3}} + 
e^{\frac{i}{2}\varphi_2} \,c_1 \,c_{12} \,s_{23} \,\sqrt{m_{\nu_2}})
\nonumber \\ 
&& 
s_2 c_3 
(
- c_1 \,c_{23}\,\sqrt{m_{\nu_3}}
+ e^{\frac{i}{2}\varphi_2} \, s_1\,c_{12} \,s_{23} \,\sqrt{m_{\nu_2}} 
) 
+ {\cal O}(\theta_{13})
\big|^2 .
\end{eqnarray} 
For $\widetilde m_{1,\mu},\widetilde m_{1,\tau}$, only the zeroth
order terms in an expansion in $\theta_{13}$ are shown.  
We note that, when compared to $\theta_2$ and $\theta_3$, 
$\theta_1$ only plays a minor role for leptogenesis. This can be
clarified by considering the second and third terms on
the right side of the equations, which contain the two potentially
large contributions associated with $\sqrt{m_{\nu_2}}$ and
$\sqrt{m_{\nu_3}}$. As can be seen, a real $\theta_1$ mainly rotates
terms proportional to $\sqrt{m_{\nu_2}}$ into terms
proportional to $\sqrt{m_{\nu_3}}$, and thus it is not introducing any
new features. 
For instance, with $\theta_1 = \pi/2$, the roles of $\theta_2$ and
$\theta_3$ are simply interchanged: a $\theta_2$ rotation now
generates contributions to $\widetilde m_{1,\alpha}$ of ${\cal
  O}(m_{\nu_2})$ and a $\theta_3$ rotation induces contributions to
$\widetilde m_{1,\alpha}$ of ${\cal O}(m_{\nu_3})$. 
Large imaginary parts of $\theta_1$ typically lead to $\widetilde
m_{1,\alpha} > m_{\nu_1} \approx m^*$, and thus are not useful in 
achieving optimal efficiencies. 
Although it is straightforward to generalise the discussion to
include arbitrary $\theta_1$, we will simplify the analysis by
setting $\theta_1 = 0$ in what follows.  
In the limit of $\theta_1 = 0$, we find
\begin{eqnarray}
\widetilde m_{1,e} &\approx&
\big| 
e^{\frac{i}{2}\varphi_1}\, c_2\,c_3\, c_{13} \, c_{12} \,\sqrt{m_{\nu_1}} 
 - e^{\frac{i}{2}\varphi_2} \, s_3 \,s_{12} \,c_{13} \,\sqrt{m_{\nu_2}}
- e^{i \delta} \,s_2 \,c_3 \,s_{13}\,\sqrt{m_{\nu_3}}
\big|^2 , \label{Eq:mt_01_e}\\
\widetilde m_{1,\mu} &\approx&
\big| 
- e^{\frac{i}{2}\varphi_1}\, c_2\,c_3\, s_{12} \, c_{23} \,\sqrt{m_{\nu_1}} 
- e^{\frac{i}{2}\varphi_2}\, s_3  \,c_{12} \,c_{23} \,\sqrt{m_{\nu_2}}
- s_2\, c_3\, s_{23}\,\sqrt{m_{\nu_3}} 
\big|^2 , \label{Eq:mt_01_m}\\
\widetilde m_{1,\tau} &\approx&
\big| 
e^{\frac{i}{2}\varphi_1}\, c_2\,c_3\, s_{12} \, s_{23} \,\sqrt{m_{\nu_1}} 
+ e^{\frac{i}{2}\varphi_2} \, s_3 \,c_{12} \,s_{23} \,\sqrt{m_{\nu_2}} 
- s_2 \, c_3 \,c_{23}\,\sqrt{m_{\nu_3}} 
\big|^2 ,\label{Eq:mt_01_t}
\end{eqnarray}
where, as in Eqs.~(\ref{Eq:mt_2}, \ref{Eq:mt_3}), 
only the zeroth order terms in an
expansion in $\theta_{13}$ are shown for $\widetilde
m_{1,\mu},\widetilde m_{1,\tau}$.  
In general, for an optimal efficiency, it is crucial to avoid 
contributions to $\widetilde m_{1,\alpha}$
of order ${\cal O}(m_{\nu_2})$ or ${\cal O}(m_{\nu_3})$.
In the flavour-independent approximation, achieving 
an optimal washout (given by the sum 
$\widetilde m_1 = \sum_\alpha \widetilde m_{1,\alpha}$)  
necessarily required that both $|s_{2}|$
and $|s_3|$ were small \cite{Antusch:2006vw}.
As can be seen from Eqs.~(\ref{Eq:mt_01_e}-\ref{Eq:mt_01_t}), when 
flavour effects are included, washout considerations still favour 
parameter regions with small $|s_{2}|$ and $|s_3|$. Nevertheless,
in the flavour-dependent treatment, the individual washout parameters
$\widetilde m_{1,\alpha}$ are typically smaller than their sum,
$\widetilde m_1$, and can significantly differ from each other. It
is therefore pertinent to re-investigate whether other regions of
parameter space may also allow for optimal washout.  
 
Let us first consider the contribution to $\widetilde m_{1,\tau}$ (or
similarly to $\widetilde m_{1,\mu}$) proportional to $m_{\nu_2}$, due
to non-zero values of $\theta_3$.  
From Eq.~(\ref{Eq:mt_01_t}) we find that this is given by 
$c_{12}^2\, s_{23}^2\, s^2_3 \,m_{\nu_2} \,\approx \,\tfrac{1}{3}\,
s^2_3 \,m_{\nu_2}$. 
In addition, the quantity which enters the efficiency factor 
is the product $A_{\tau\tau}  \widetilde m_{1,\tau}$, 
with $A_{\tau\tau}\approx -0.6$. 
Combining these two effects reduces the $\theta_3$-induced washout 
by a factor of about $1/5$, and even for $s_3 \approx {\cal O}(1)$  
optimal washout $|A_{\alpha\alpha}| \widetilde m_{1,\alpha} \approx m^*$
is still possible to obtain. 
On the other hand, the contribution to $\widetilde m_{1,\tau}$ (or
similarly to $\widetilde m_{1,\mu}$) proportional to $m_{\nu_3}$ due
to non-zero $\theta_2$ 
is given by $ c_{23}^2\, s^2_2 \,c^2_3\,m_{\nu_3} \,\approx \,
\tfrac{1}{2}\, s^2_3 \,c^2_3\,m_{\nu_3}$. 
Again, the quantity which enters the efficiency factor is the product
$|A_{\tau\tau}| \widetilde m_{1,\tau}$. Although combining these two
effects reduces the $\theta_2$-induced washout by a factor of about $0.3$,
optimal washout cannot be achieved for large $|s_2|$, if $\theta_3$
is small. The only exception occurs for small $|c_3|$, which can
suppress the large contribution of ${\cal O}(m_{\nu_3})$ to
$\widetilde m_{1,\tau},\widetilde m_{1,\mu}$ and still allow for large
$|s_2|$. However, as we will see in the next subsection, the decay
asymmetries $\varepsilon_{1,\alpha}$ are somewhat suppressed in this
case.

Another difference between the flavour-independent approximation and the
correct flavour-dependent treatment becomes apparent when we consider
$\widetilde m_{1,e}$. 
In contrast to $\widetilde m_{1,\tau}$ and $\widetilde m_{1,\mu}$,
large values of $s_2$ only induce a washout parameter $\widetilde m_{1,e}$
of $s^2_{13} \times {\cal O}(m_{\nu_3})$.
In other words, we can be in the optimal 
washout regime $\widetilde m_{1,e} \approx
m^*$ also for larger values of $s_2$, depending on the size of
$\theta_{13}$. Nevertheless, 
in this case the decay asymmetry $\varepsilon_{1,e}$ is also 
suppressed when compared to the optimal regime. 

Let us point out that another possible way to obtain optimal $\widetilde
m_{1,\mu}$ and $\widetilde m_{1,\tau}$ of order $m^*$ would be to
align $s_2$ and $s_3$ such that their contributions to one of the 
washout parameters (respectively proportional to $m_{\nu_3}$ and $m_{\nu_2}$)
would nearly cancel each other. In this case one could obtain, for example,  
$\widetilde m_{1,\mu} \approx m^*$.
However, one can verify that the third washout parameter 
$\widetilde m_{1,\tau}$ (and thus $\widetilde m_1$) 
would still be ${\cal O}(m_{\nu_3})$ or
${\cal O}(m_{\nu_2})$, which would imply a suppression of 
$\varepsilon_{1,\mu}$ by a factor $m^*/\widetilde m_{1}$,
in comparison to $\varepsilon_1^\mathrm{max}$
(c.f.~Eq.~(\ref{Eq:BoundOnEpsa})).

\subsubsection{Flavour-Dependent Decay
 Asymmetries}\label{Sec:FlDecayAsymmetries} 

From the analysis of the flavour-dependent washout parameters 
$\widetilde m_{1,\alpha}$, it has become apparent that there are several  
regions of the seesaw parameter space with appealing 
prospects for leptogenesis. They can be summarised as follows.
Generically favoured is the region of small $|s_2|$ and $|s_3|$, 
where the washout parameters $\widetilde m_{1,\alpha}$ are 
${\cal O}(m_{\nu_1})$, and receive only small contributions
proportional to $m_{\nu_2}$ and $m_{\nu_3}$. 
In the flavour-independent approximation, this was the only  
region favoured by washout~\cite{Antusch:2006vw}.  
In contrast, in flavour-dependent leptogenesis,
washout (in all flavours) remains optimal also for larger values of 
$|s_3| \lesssim 1$. 
With small $|c_3|$ (implying large $|s_3|$ which is no longer
disfavoured), large $|s_2|$ becomes compatible with optimal washout, 
and a whole new region, washout-favoured, has emerged. 
An additional interesting effect is that washout in the {\small$e$}-flavour, 
governed by $\widetilde m_{1,e}$, can also be optimal for large
$|s_2|$, due  to the smallness of $\theta_{13}$.

In addition to an optimal $\widetilde m_{1,\alpha}$, and
given the tight constraints on the reheat temperature, it is also
desirable to have decay asymmetries close to the optimal value
$\varepsilon_1^\mathrm{max}$. Let us now address the latter issue more
thoroughly. 
First, we note that for $R=\mathbbm{1}$, and more generally,  
for $\theta_i = 0,\tfrac{\pi}{2}$ (mod $\pi$), the decay asymmetries
exactly vanish. A deviation from the latter values of the form 
$\theta_2 = \theta_3 = 0$ but $\theta_1 \not= 0$, also leads to zero
values.
To clarify the analysis, let us explicitly write the decay
asymmetries in terms of the seesaw parameters. For simplicity, and 
as illustrative examples, we first consider the dependence on $\theta_2$,
with $\theta_{3} = \theta_{1} = 0$, and then study the $\theta_3$-dependence, 
setting $\theta_{2} = \theta_{1} = 0$. 
To discuss the parameter regions with optimal washout where $\theta_2$ 
and $\theta_3$ are both large, we then turn to the decay 
asymmetries with nonzero $\theta_2$ and $\theta_3$.
For simplicity, in the latter case we will again present the formulae 
with $\theta_{1} = 0$. 
The discussion can be easily generalised to arbitrary $\theta_{i}$ values,
using the general expressions for the decay asymmetries as given in
Eqs.~(\ref{Eq:EpsMSSM}, \ref{Eq:Epsa_R}). 
We will also illustrate the effects of nonzero $\theta_1$ via numerical
examples in Section~\ref{Sec:NumericalExamples}. 

For $\theta_{3} = \theta_{1} = 0$, one obtains
\begin{eqnarray}\label{Eq:eps_e_t2}
\varepsilon_{1,e} \!&\approx&\!
-\frac{3\,m_{N_1}}{8\,\pi \,v_2^2\,(|c_2|^2\,{m_{\nu_1}} +
|s_2|^2\,{m_{\nu_3}} )}\,
\mathrm{Im}\left[
\,{s^2_{13}}\,{s^2_2}\, m_{\nu_3}^2\,\right. \nonumber \\
&& \left.- \,
e^{\frac{\imag}{2}( {{\varphi}_1}-2 \,\delta )}\,
{s_{13}}\, {c_{13}}\,{c_{12}}\,{s_2}\,{c_2}\,
{\sqrt{m_{\nu_1}}} \,m_{\nu_3}^{\frac{3}{2}} + \dots
\right]
\, ,\\
\label{Eq:eps_m_t2}
\varepsilon_{1,\mu} \!&\approx&\!
-\frac{3\,m_{N_1}}{8\,\pi \,v_2^2\,(|c_2|^2\,{m_{\nu_1}} +
|s_2|^2\,{m_{\nu_3}} )}\,
\mathrm{Im}\left[
\,{s^2_{23}}\,{s^2_2}\,{c^2_{13}}\, m_{\nu_3}^2\,
\right. \nonumber \\
&& \left.+\,
e^{\frac{\imag}{2}{{\varphi}_1}}{c_{13}}\,\,{s_{12}}\,{s_{23}}\,
{c_{23}}\,{s_2}\,{c_2}\, {\sqrt{m_{\nu_1}}}
\,m_{\nu_3}^{\frac{3}{2}}\right. \nonumber \\
&& \left.
+ \,
e^{\frac{\imag}{2}({{\varphi}_1}-2\delta)}\,
{s_{13}}\,{c_{13}}\,{c_{12}}\,{s^2_{23}}\,{s_2}\,{c_2}\, {\sqrt{m_{\nu_1}}}
\,m_{\nu_3}^{\frac{3}{2}}
+ \dots
\right]
\, ,\\
\label{Eq:eps_t_t2}
\varepsilon_{1,\tau} \!&\approx&\!
-\frac{3\,m_{N_1}}{8\,\pi \,v_2^2\,(|c_2|^2\,{m_{\nu_1}} +
|s_2|^2\,{m_{\nu_3}} )}\,
\mathrm{Im}\left[
\,{c^2_{23}}\,{s^2_2}\,{c^2_{13}}\, m_{\nu_3}^2\,\right. \nonumber \\
&& \left.-\,
e^{\frac{\imag}{2}{{\varphi}_1}}\,{c_{13}}\,{s_{12}}\,{s_{23}}\,
{c_{23}}\,{s_2}\,{c_2}\, {\sqrt{m_{\nu_1}}} \,m_{\nu_3}^{\frac{3}{2}}
\right. \nonumber \\
&& \left.
+ \,
e^{\frac{\imag}{2}({{\varphi}_1}-2\delta)}
\,{s_{13}}\,{c_{13}}\,{c_{12}}\,{c^2_{23}}\,
{s_2}\,{c_2}\, {\sqrt{m_{\nu_1}}} \,m_{\nu_3}^{\frac{3}{2}}
+ \dots
\right] \, ,
\end{eqnarray}
where the dots indicate terms
${\cal O}(m_{\nu_1}^{\frac{3}{2}}\, \sqrt{m_{\nu_3}})$
and ${\cal O}(m_{\nu_1}^2)$, which are sub-leading for hierarchical light
neutrinos.
Eqs.~(\ref{Eq:eps_m_t2}, \ref{Eq:eps_t_t2}) show that, 
provided that $\theta_2$ has an imaginary part, the 
$\theta_2$-induced decay asymmetries $\varepsilon_{1,\mu}$ and
$\varepsilon_{1,\tau}$ can be close to the optimal value
$\varepsilon_1^\mathrm{max}$ defined in Eq.~(\ref{Eq:BoundOnEpsa}). 
Regarding the sub-leading terms in 
Eqs.~(\ref{Eq:eps_m_t2}, \ref{Eq:eps_t_t2}), decay asymmetries 
can also emerge if the Majorana CP phase $\varphi_1$ is nonzero,
albeit suppressed 
by a factor ${\cal O}(\sqrt{m_{\nu_1}/m_{\nu_3}})$.
On the other hand, 
for hierarchical light neutrinos, the leading contributions
to $\varepsilon_{1,e}$
are suppressed by $s^2_{13}$ and $s_{13}$ when compared to
$\varepsilon_{1,\mu}$ and $\varepsilon_{1,\tau}$.
This is in agreement with Eq.~(\ref{Eq:BoundOnEpsa}), which states
that for large $|s_2|$ (and thus $\widetilde m_1 
\approx {\cal O}(m_{\nu_3})$) the decay asymmetry is
suppressed by a factor $\sqrt{\widetilde m_{1,\alpha}/\widetilde
m_1} \approx  s_{13}$, compared to the maximal value,
$\varepsilon_1^\mathrm{max}$. 

Let us now consider the case where $\theta_{2} = \theta_{1} = 0$. The
flavour-dependent decay asymmetries are as follows:
\begin{eqnarray}
\varepsilon_{1,e} \!\!&\approx&\!\!
-\frac{3\,m_{N_1}}{8\,\pi \,v_2^2\,(|c_3|^2\,{m_{\nu_1}} +
|s_3|^2\,{m_{\nu_2}} )}\,
\mathrm{Im}\left[
\,{s^2_{12}}\,{c^2_{13}}\,{s^2_3}\, m_{\nu_2}^2\,\right.\nonumber \\
&& \left.
-\,
e^{\frac{\imag}{2}( {{\varphi}_1} - {{\varphi}_2})}\, {c^2_{13}}\,
{s_{12}}\,{c_{12}}\,{s_3}\,{c_3}\, {\sqrt{{m_{\nu_1}}}} \,
m_{\nu_2}^{\frac{3}{2}}
+ \dots
\right] \,, \label{Eq:eps_e_t3} \\
\varepsilon_{1,\mu} \!\!&\approx&\!\!
-\frac{3\,m_{N_1}}{8\,\pi \,v_2^2\,(|c_3|^2\,{m_{\nu_1}} +
|s_3|^2\,{m_{\nu_2}} )}\,
\mathrm{Im}\left[
\,({c^2_{12}}\,{c^2_{23}} + {s^2_{13}}\, {s^2_{12}}\,{s^2_{23}} )\,{s^2_3}\,
m_{\nu_2}^2\,\right. \nonumber \\
&& \left. + \,
e^{\frac{\imag}{2}( {{\varphi}_1} -  {{\varphi}_2})} \,{s_3}\,{c_3}
(c_{23} \,{s_{12}} + e^{-\imag \delta}  {s_{13}}\,s_{23} \,{c_{12}} )( c_{23}
\,{c_{12}} -  e^{\imag \delta} {s_{13}} s_{23} \,{s_{12}})
{\sqrt{{m_{\nu_1}}}} \,m_{\nu_2}^{\frac{3}{2}} \,\right. \nonumber \\
&& \left.+ \, \dots
\right] \, ,   \label{Eq:eps_m_t3} \\
\varepsilon_{1,\tau} \!\!&\approx&\!\!
-\frac{3\,m_{N_1}}{8\,\pi \,v_2^2\,(|c_3|^2\,{m_{\nu_1}} +
|s_3|^2\,{m_{\nu_2}} )}\,
\mathrm{Im}\left[
\,({c^2_{12}}\,{s^2_{23}}  + {s^2_{13}}\,{s^2_{12}}\,{c^2_{23}})\,{s^2_3}\,
m_{\nu_2}^2\, \right.\nonumber \\
&& \left. + \,
e^{\frac{\imag}{2}( {{\varphi}_1} - {{\varphi}_2})}\, {s_3}\,{c_3}\,
(s_{23} \,{s_{12}} - e^{-\imag \delta}  {s_{13}}\,c_{23} \,{c_{12}} )( s_{23}
\,{c_{12}} +  e^{\imag \delta} {s_{13}} c_{23} \,{s_{12}})
{\sqrt{{m_{\nu_1}}}} \,m_{\nu_2}^{\frac{3}{2}} \,\right. \nonumber \\
&& \left.+ \, \dots
\right] \,  .\label{Eq:eps_t_t3}
\end{eqnarray}
In the above, the dots denote sub-leading terms
${\cal O}(m_{\nu_1}^{\frac{3}{2}}\, \sqrt{m_{\nu_3}})$,
${\cal O}(m_{\nu_1}^2)$.
Compared to $\varepsilon_1^\mathrm{max}$, the decay asymmetries
induced by $\theta_3$ are suppressed by a factor of
${\cal O}(m_{\nu_2}/m_{\nu_3}) \approx 1/5$ with respect to those
induced by $\theta_2$.
As can be seen from Eqs.~(\ref{Eq:eps_e_t3}-\ref{Eq:eps_t_t3}),
either a non-zero imaginary part of $\theta_3$ or non-zero phase
$\delta$ or the difference $\varphi_1 - \varphi_2$, 
can in principle provide the required CP violation for leptogenesis.

In our discussion of the seesaw parameters with optimal washout, we have 
encountered a new region where $|s_3|$ and $|s_2|$ were large, but 
with small $|c_3|$. In the limit of vanishing $\theta_1$, 
and keeping only the leading terms for simplicity, we obtain
\begin{eqnarray}
\label{Eq:eps_e_t2_t3}
\varepsilon_{1,e} \!\!&\approx&\!\!
-\frac{3\,m_{N_1}}{8\,\pi \,v_2^2\,(|c_2 c_3|^2\,{m_{\nu_1}} +
|s_3|^2\,{m_{\nu_2}} + |s_2 c_3|^2\,{m_{\nu_3}})}\,
\mathrm{Im}\left[
\,{s^2_2}\,{c^2_3}\,{s^2_{13}}\, m_{\nu_3}^2\,\right. \\
&& \left.
+\,
e^{\frac{\imag}{2}( {{\varphi}_2} - 2 \delta)}\, {s_{13}}\,{c_{13}}\,
{s_{12}}\,{s_{2}}\,{s_3}\,{c_3}\, {\sqrt{{m_{\nu_2}}}} \,
m_{\nu_3}^{\frac{3}{2}}
+ \dots
\right] \,,\nonumber \\
\label{Eq:eps_m_t2_t3}
\varepsilon_{1,\mu} \!\!&\approx&\!\!
-\frac{3\,m_{N_1}}{8\,\pi \,v_2^2\,(|c_2 c_3|^2\,{m_{\nu_1}} +
|s_3|^2\,{m_{\nu_2}} + |s_2 c_3|^2\,{m_{\nu_3}})}\,
\mathrm{Im}\left[
\,{s^2_2}\,{c^2_3}\,{s^2_{23}}\,m_{\nu_3}^2\,\right.  \\
&& \left. + \,
e^{\frac{\imag}{2} {{\varphi}_2}}\,{c_{12}}\,{s_{23}}\,{c_{23}}
\,{s_{2}}\,{s_3}\,{c_3}\, {\sqrt{{m_{\nu_2}}}} \,
m_{\nu_3}^{\frac{3}{2}} + {\cal O}(s_{13}) \, {\sqrt{{m_{\nu_2}}}} \,
m_{\nu_3}^{\frac{3}{2}} + \, \dots
\right] \, ,  \nonumber \\
\label{Eq:eps_t_t2_t3}
\varepsilon_{1,\tau} \!\!&\approx&\!\!
-\frac{3\,m_{N_1}}{8\,\pi \,v_2^2\,(|c_2 c_3|^2\,{m_{\nu_1}} +
|s_3|^2\,{m_{\nu_2}} + |s_2 c_3|^2\,{m_{\nu_3}})}\,
\mathrm{Im}\left[
\,{s^2_2}\,{c^2_3}\,{c^2_{23}}\,m_{\nu_3}^2\, \right. \\
&& \left. + \,
e^{\frac{\imag}{2} {{\varphi}_2}}\,{c_{12}}\,{s_{23}}\,{c_{23}}
\,{s_{2}}\,{s_3}\,{c_3}\, {\sqrt{{m_{\nu_2}}}} \,
m_{\nu_3}^{\frac{3}{2}} + {\cal O}(s_{13}) \, {\sqrt{{m_{\nu_2}}}} \,
m_{\nu_3}^{\frac{3}{2}} + \, \dots
\right] \nonumber . 
\end{eqnarray}
In the above, the dots denote sub-leading terms
${\cal O}(m_{\nu_1}^{\frac{3}{2}}\, \sqrt{m_{\nu_3}})$, ${\cal
  O}(m_{\nu_2}^{\frac{3}{2}}\, \sqrt{m_{\nu_3}})$, 
${\cal O}(m_{\nu_1}^2)$ and ${\cal O}(m_{\nu_2}^2)$.
Eqs.~(\ref{Eq:eps_e_t2_t3}-\ref{Eq:eps_t_t2_t3}) show that 
for large $|s_3|$, the asymmetries are suppressed 
by $|c_3|/|s_3|$ when compared to the 
optimal value $\varepsilon_1^\mathrm{max}$ defined in 
Eq.~(\ref{Eq:BoundOnEpsa}). 
Although it is possible to achieve optimal washout in the region with
large $|s_3|$ and $|s_2|$ if $|c_3|$ is small, 
the smallness of $|c_3|$ in turn leads to somewhat suppressed decay
asymmetries in all flavours. 

To conclude this discussion, 
let us stress that as seen in the above considered cases
(c.f. Eqs.~(\ref{Eq:eps_e_t2}-\ref{Eq:eps_t_t2_t3})), and contrary to
what occurred in the flavour-independent approximation, even for a real
$R$-matrix, one can indeed obtain non-vanishing values for the baryon
asymmetry. Notice however that the contributions of the $U_\text{MNS}$ 
phases appear suppressed by ratios of the light neutrino masses 
($\sqrt{m_{\nu_i}/m_{\nu_j}}$, {\small$i < j$}) and/or by $\theta_{13}$.
Also notice that other new regions of the parameter space 
where $|s_2|$ and $|s_3|$ can be large (with complex
$R$) have good prospects for leptogenesis, but
lead to decay asymmetries which are suppressed when 
compared to the favoured region of small $|s_2|$ and $|s_3|$.

In the following subsection, we will discuss whether or not the promising
regions here identified still offer viable BAU scenarios when bounds on
$T_\text{RH}$ (namely $T_\mathrm{RH} \le 2 \times 10^{10}$ GeV and 
$T_\mathrm{RH} \le 5 \times 10^{9}$ GeV) are taken into account.

\subsection{Numerical examples}\label{Sec:NumericalExamples}

The analysis of the previous subsections, based on the analytical
expressions for the flavour dependent washout parameters $\widetilde
m_{1,\alpha}$ and decay asymmetries $\varepsilon_{1,\alpha}$,
has revealed interesting differences between the constraints on the
seesaw parameter space arising from leptogenesis in the
flavour-independent approximation and the correct flavour-dependent 
treatment.
In what follows, we will present some numerical examples which
illustrate these new constraints, taking into account bounds
from $T_\text{RH}$. In all the following examples, we will always
consider $m_{\nu_1}=10^{-3}$ eV.

Before beginning, let us notice that the figures here displayed show ranges
of the maximal attainable BAU. As done before, a scan is performed
over $m_{N_1}$, and its value is determined as to obtain maximal 
$n_\text{B}/n_\gamma$. 
Additionally, it is important to stress that, in agreement with the
discussion of Section~\ref{Sec:limitations}, there may still exist
significant theoretical uncertainties in the estimates of the produced
baryon asymmetry. As previously mentioned, the effect of these
uncertainties is hard to quantify, and can lead to both over- and
under-estimations of the BAU. An educated guess of these
theoretical uncertainties would suggest that one should allow for as
much as a factor 2 (or even 5) between the real and the estimated
values.
Thus, when evaluating the BAU viability of the seesaw parameter space,
we will also be showing regions where the
produced baryon-to-photon ratio lies outside the WMAP observed range, 
but is still larger than $10^{-10}$. In particular, we allow for a factor
2 (5) arising from theoretical uncertainties, and display the
corresponding regions $n_\text{B}/n_\gamma 
\in [3 \times 10^{-10}, \,5.9 \times 10^{-10}]$ ($n_\text{B}/n_\gamma 
\in [10^{-10}, \,3 \times 10^{-10}]$) in green (grey).
If our computations were exact, the shown
region with $n_\text{B}/n_\gamma \ge 5.9 \times 10^{-10}$ could be
compatible with the observed baryon asymmetry (notice that values
larger than the WMAP range can be easily accommodated by varying
$m_{N_1}$).  

Even in the absence of considering the new CP violation sources
arising from the $U_\text{MNS}$ matrix, the flavour-dependent
computation gives rise to interesting 
new constraints on the seesaw parameter space.
Thus, we first examine a conservative scenario with CP violation exclusively
stemming from complex $R$-matrix angles, taking into account the
bounds for the reheat temperature.     

\begin{figure}
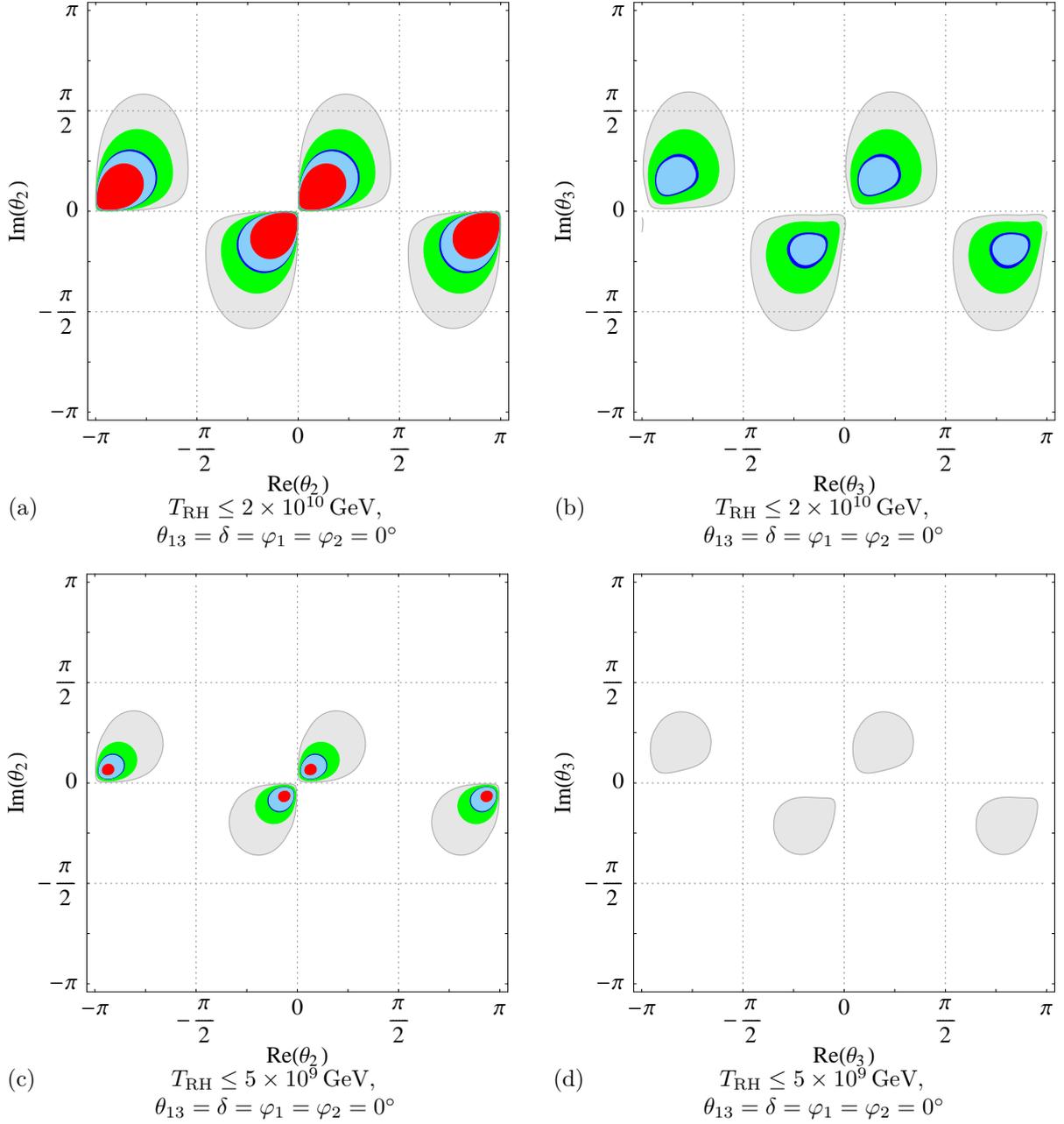

 \centering   
 \subfigure[$\quad\quad 
 T_{\mathrm{RH}} \le 2 \times 10^{10}\,\text{GeV},\,\quad\quad $
$ \quad \quad \theta_{13}=\delta=\varphi_1=\varphi_2=0^\circ
   $]{$\CenterEps[0.895]{plot_t13_0_d_0_Re_t2_Im_t2_ScanN}$} 
 \;\;\;\, 
 \subfigure[$\quad\quad T_{\mathrm{RH}} \le 2 \times
   10^{10}\, \text{GeV},\,\quad\quad$
   $\theta_{13}=\delta=\varphi_1=\varphi_2= 0^\circ
   $]{$\CenterEps[0.895]{plot_t13_0_d_0_Re_t3_Im_t3_ScanN}$} 
 \\
 \subfigure[$\quad\quad T_{\mathrm{RH}} \le 5 \times
 10^{9}\, \text{GeV},\,\quad\quad $
 $\theta_{13}=\delta=\varphi_1=\varphi_2= 0^\circ
 $]{$\CenterEps[0.895]{plot_t13_0_d_0_Re_t2_Im_t2_ScanN_5_9}$} 
 \;\;\;\, 
 \subfigure[$\quad\quad T_{\mathrm{RH}} \le 5 \times
 10^{9}\, \text{GeV},\,\quad\quad $
 $\theta_{13}=\delta=\varphi_1=\varphi_2= 0^\circ
 $]{$\CenterEps[0.895]{plot_t13_0_d_0_Re_t3_Im_t3_ScanN_5_9}$} 
 \caption{\label{fig:ReImt2t3}
Regions of $\mathrm{Re}(\theta_{2})-\mathrm{Im}(\theta_{2})$ and
$\mathrm{Re}(\theta_{3})-\mathrm{Im}(\theta_{3})$ parameter spaces
compatible with successful thermal leptogenesis in the presence of 
bounds on the reheat temperature:
$T_{\mathrm{RH}} \le 2 \times 10^{10}$ GeV (upper panels) and 
$T_{\mathrm{RH}} \le 5 \times 10^{9}$ GeV (lower panels).
From out- to inner-most regions, 
the associated ranges of maximally possible baryon asymmetry are:
$n_\text{B}/n_\gamma \in [10^{-10}, \,3 \times 10^{-10}]$,
$n_\text{B}/n_\gamma \in [3 \times 10^{-10}, \,5.9 \times 10^{-10}]$,
$n_\text{B}/n_\gamma
\in [5.9 \times 10^{-10}, \,6.3 \times 10^{-10}]$, $n_\text{B}/n_\gamma
\in [6.3 \times 10^{-10}, \,10^{-9}]$ and 
$n_\text{B}/n_\gamma \gtrsim 10^{-9}$. 
The corresponding colour code is grey, green, dark blue (WMAP), 
light blue and red. }
\end{figure}

Figure~\ref{fig:ReImt2t3} illustrates the $\mathrm{Re}(\theta_{2}) -
\mathrm{Im}(\theta_{2})$ and $\mathrm{Re}(\theta_{3}) -
\mathrm{Im}(\theta_{3})$ regions compatible with successful thermal
leptogenesis in the presence of bounds $T_{\mathrm{RH}} \le 2 \times
10^{10}$ GeV and $T_{\mathrm{RH}} \le 5 \times 10^{9}$ GeV. In this case
$\theta_{13},\delta,\varphi_1,\varphi_2$ and $\theta_{1}$ 
have been set to zero. On the left (right) panels, $\theta_3=0$
($\theta_2=0$). 
The examples with $T_{\mathrm{RH}} \le 2 \times 10^{10}$ GeV (i.e. 
Fig.~\ref{fig:ReImt2t3}(a) and Fig.~\ref{fig:ReImt2t3}(b))
update the analysis of
Ref.~\cite{Antusch:2006vw}, which had been performed in the
flavour-independent approximation. In the present flavour-dependent
computation, we find that the BAU arising from complex 
$\theta_{3}$ is somewhat larger and new regions, which are indeed
compatible with the WMAP range, have now emerged.
It is worth stressing that in this case of complex $R$-matrix
angles, the favoured regions still correspond to small values of $\theta_2$
and $\theta_3$. 
Considering a stronger bound on $T_\text{RH}$, namely 
$T_{\mathrm{RH}} \le 5 \times 10^{9}$ GeV, we notice that there are
still regions in the $\mathrm{Re}(\theta_{2}) -
\mathrm{Im}(\theta_{2})$ plane compatible with WMAP 
observations. 
When the latter bound on $T_\text{RH}$ is applied,
we verify that for complex
$\theta_3$ values it is no longer possible to saturate
the WMAP preferred range. Nevertheless, regions where
$n_\text{B}/n_\gamma \in [10^{-10}, \,3 \times 10^{-10}]$ can still
be found (viable if one allowed for a factor 5 uncertainty in the
computation). In any case, it is manifest that for this stricter
$T_\text{RH}$ bound, the preferred source of CP violation for
leptogenesis is $\theta_2$.
In both cases, the observed differences between the present and the
previous analyses (\cite{Antusch:2006vw}) originate from taking
into account flavour effects in the Boltzmann equations. In the
$\theta_2$ case, the differences are less apparent, essentially due to
the fact that $\widetilde m_{1,e} = 0$ and $\widetilde m_{1,\mu} \approx
\widetilde m_{1,\tau}$. 
However, important effects can be observed for the $\mathrm{Re}(\theta_{3}) -
\mathrm{Im}(\theta_{3})$ plane, since in this case both the decay
asymmetries and washout parameters differ for each individual
flavour. In particular, this leads to deformations of the allowed
regions when compared to those presented in Ref.~\cite{Antusch:2006vw}.  

\begin{figure}
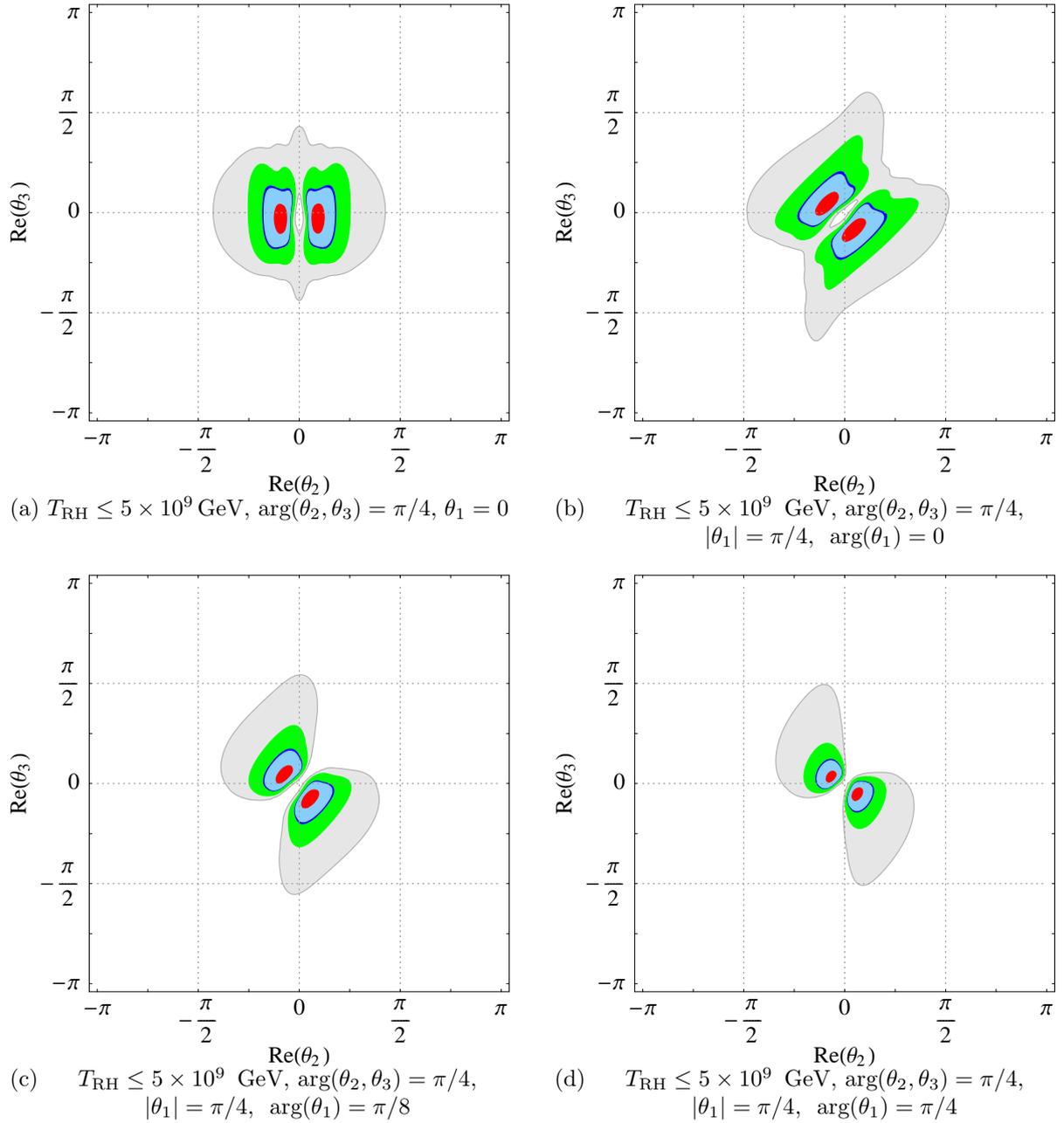

 \centering   
 \subfigure[
   $ T_{\mathrm{RH}} \le 5 \times 10^{9}\,
 \text{GeV},\,\mbox{arg}(\theta_2,\theta_3)= \pi/4,$ $\theta_1=0 $
 ]{$\CenterEps[0.895]{plot_t13_0_d_0_ArgPio4_t1_0_Re_t2_Re_t3_ScanN}$}  
 \;\;\;\, 
 \subfigure[
 $
 T_{\mathrm{RH}}\le 5\times 10^{9}\,~\text{GeV},\,
 \mbox{arg}(\theta_2,\theta_3)=\pi/4,$ 
 $|\theta_1|=\pi/4,\,~\mbox{arg}(\theta_1)=0 $ 
 ]{$\CenterEps[0.895]{plot_t13_0_d_0_ArgPio4_t1_Pio4_Re_t2_Re_t3_ScanN}$}  
 \\ 
\subfigure[
 $ T_{\mathrm{RH}}\le 5 \times 10^{9}\,~\text{GeV},\,
 \mbox{arg}(\theta_2,\theta_3)=\pi/4,$
 $|\theta_1|=\pi/4,\,~\mbox{arg}(\theta_1)=\pi/8 $  
]{$\CenterEps[0.895]{plot_t13_0_d_0_ArgPio4_t1_Pio4_Argt1_Pio8_Re_t2_Re_t3_ScanN}$}   
 \;\;\;\, 
 \subfigure[
 $ T_{\mathrm{RH}} \le 5 \times 10^{9}\,~\text{GeV},\,
 \mbox{arg}(\theta_2,\theta_3)=\pi/4,$
 $|\theta_1|=\pi/4,\,~\mbox{arg}(\theta_1)=\pi/4$   
 ]{$\CenterEps[0.895]{plot_t13_0_d_0_ArgPio4_t1_Pio4_Argt1_Pio4_Re_t2_Re_t3_ScanN}$}  
 \caption{\label{fig:ReImt2t3wt1} 
Regions of $\mathrm{Re}(\theta_{2})-\mathrm{Re}(\theta_{3})$ parameter space
compatible with successful thermal leptogenesis in the presence of a
bound $T_{\mathrm{RH}} \le 5 \times 10^{9}$ GeV. 
We display several values of complex $\theta_1$, choosing
$\mbox{arg}(\theta_2)=\mbox{arg}(\theta_3)=\pi/4$ and 
$\theta_{13}=\delta=\varphi_1=\varphi_2=0^\circ$. 
Colour code as in Fig.~\ref{fig:ReImt2t3}. }
\end{figure}

In Fig.~\ref{fig:ReImt2t3wt1} we illustrate the effects of non-zero
$\theta_1$.  Taking $\mbox{arg}(\theta_2)=\mbox{arg}(\theta_3)=\pi/4$,
and again $\theta_{13},\delta,\varphi_1,\varphi_2=0^\circ$, we now display
the regions of the $\mathrm{Re}(\theta_{2}) - \mathrm{Re}(\theta_{3})$
parameter space compatible with thermal leptogenesis in the presence
of a bound $T_{\mathrm{RH}} \le 5 \times 10^{9}$ GeV.  
First, let us point out that Fig.~\ref{fig:ReImt2t3wt1}(a) corresponds
to a variation of Figs.~\ref{fig:ReImt2t3}(c) and (d), but for fixed values of 
$\mbox{arg}(\theta_2)$ and $\mbox{arg}(\theta_3)$.
When compared to Fig.~\ref{fig:ReImt2t3wt1}(a),
Fig.~\ref{fig:ReImt2t3wt1}(b) shows the effect of $\theta_1=\pi/4$,
which is mainly a rotation (and a slight deformation) of the
allowed region. Figure~\ref{fig:ReImt2t3wt1}(c) and 
Fig.~\ref{fig:ReImt2t3wt1}(d) illustrate that
an imaginary part of $\theta_1$, in addition to introducing an
additional source of CP violation, leads to a reduction of the
compatible parameter space. 
As argued in Section~\ref{Sec:Washout}, this effect can be explained by
a stronger washout due to an enhancement of the parameters $\widetilde
m_{1,\alpha}$. We emphasise that for complex $\theta_i$ (but
vanishing low-energy CP phases), even in the presence of
non-zero values of $\theta_1$, small values of $|s_2|$ and $|s_3|$ are still
favoured by thermal leptogenesis.

\begin{figure}
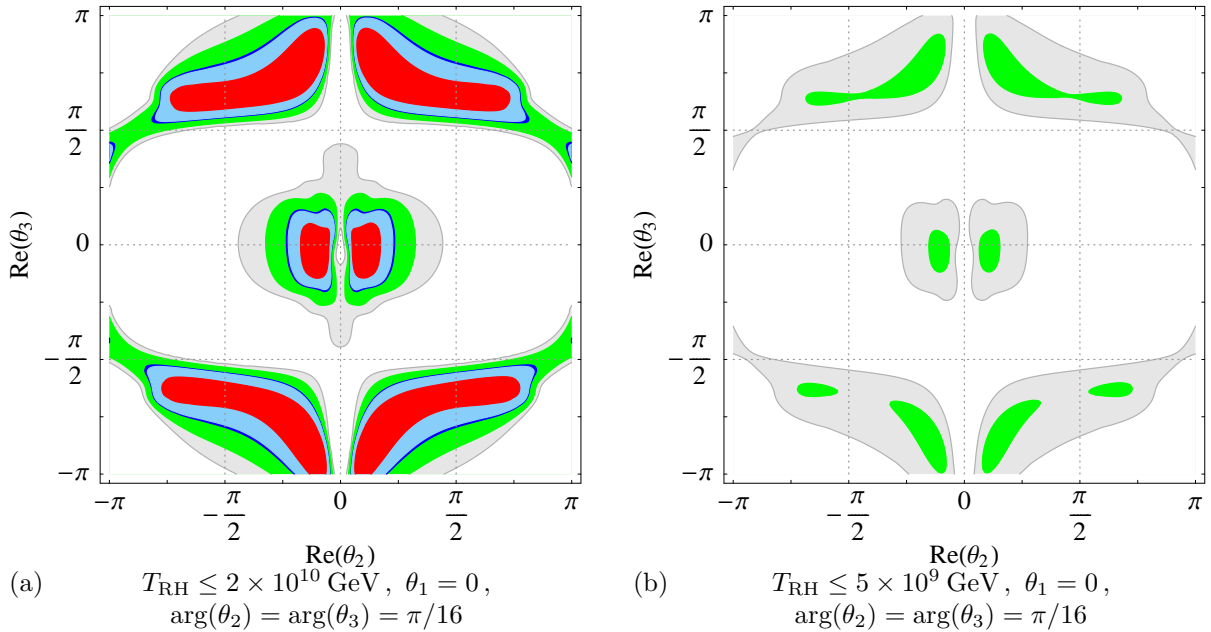

 \centering   
 \subfigure[$  T_{\mathrm{RH}} \le 2 \times
 10^{10}\:\mbox{GeV}\,,\,\,\theta_1=0\,,$ $\,\mbox{arg}(\theta_2) = \mbox{arg}(\theta_3) =
 \pi/16$
 ]{$\CenterEps[0.895]{plot_t13_0_Argt2_Pio16_Argt3_Pio16_Re_t2_Re_t3_ScanN}$} 
 \;\;\;\, 
 \subfigure[$T_{\mathrm{RH}} \le 5 \times
 10^{9}\:\mbox{GeV}\,,\,\,\theta_1=0\,,$ $\,\mbox{arg}(\theta_2) = \mbox{arg}(\theta_3) =
 \pi/16$
 ]{$\CenterEps[0.895]{plot_t13_0_Argt2_Pio16_Argt3_Pio16_Re_t2_Re_t3_ScanN_5_9}$}
 \caption{\label{fig:Arg:0MNS}
Regions of $\mathrm{Re}(\theta_{2})-\mathrm{Re}(\theta_{3})$
compatible with successful thermal leptogenesis. We take
$\theta_{1}=0$, and consider 
$\arg (\theta_2) = \arg (\theta_3) = \pi/16$.
$T_{\mathrm{RH}} \le 2 \times 10^{10}$ GeV in (a),
while $T_{\mathrm{RH}} \le 5 \times 10^{9}$ GeV in (b).
In this example we have set 
$\theta_{13}=\delta=\varphi_1=\varphi_2=0^\circ$.
Colour code as in Fig.~\ref{fig:ReImt2t3}.  }
\end{figure}

In Fig.~\ref{fig:ReImt2t3} we have separately considered the effects 
of each of the $R$-matrix angles $\theta_{2,3}$, 
while in Fig.~\ref{fig:ReImt2t3wt1}
we analysed the impact of non-vanish $\theta_1$ upon the 
$\mathrm{Re}(\theta_{2})-\mathrm{Re}(\theta_{3})$ parameter space,
assuming sizable arguments for both $\theta_2$ and $\theta_3$.
However, for quite small values of the arguments, and when flavour effects are
taken into account, new interesting regions
of the $\mathrm{Re}(\theta_{2})-\mathrm{Re}(\theta_{3})$ parameter space can
also arise. 
This is shown in Fig.~\ref{fig:Arg:0MNS}, where 
we now display the regions of the
$\mathrm{Re}(\theta_{2})-\mathrm{Re}(\theta_{3})$ parameter space 
compatible with successful thermal leptogenesis, assuming a small
value for both arguments, namely $\pi/16$.  
As can be seen, not only do we encounter large values of BAU
associated with small values of $|s_2|$ and $|s_3|$, but new extensive
regions, with larger values of $\theta_2$
and $\theta_3$, are now present. 
The origin of these new regions exhibiting a sizable BAU
can be easily understood (in the limit of vanishing $\theta_1$) 
from the analytical considerations of 
Sections~\ref{Sec:Washout} and~\ref{Sec:FlDecayAsymmetries}. 
On the one hand, Eqs.~(\ref{Eq:mt_01_e}-\ref{Eq:mt_01_t}) show that
optimal washout is possible in two cases:
for large $s_3$, or then for large $s_3$ and $s_2$, provided that 
$c_3$ is small. 
On the other hand, from 
Eqs.~(\ref{Eq:eps_e_t2_t3}-\ref{Eq:eps_t_t2_t3}) we have seen 
that optimal decay asymmetries require contributions from 
non-zero $\theta_2$, suppressed if $c_3$ is small. 
The shape of the extensive
regions in Fig.~\ref{fig:Arg:0MNS}(a) with sizable BAU reflects the 
balance between having a sufficiently small washout, 
while at the same time succeeding in obtaining
an important decay asymmetry. 
As expected, taking into account stricter bounds on $T_\text{RH}$
leads to the disappearance of the WMAP compatible regions
(c.f.~Fig.~\ref{fig:Arg:0MNS}(b)). Nevertheless, 
regions where the BAU is close to the observed range still survive,
both for small and large $\theta_2$ and $\theta_3$. 
For non-vanishing values of $\theta_1$, just like discussed regarding
Fig.~\ref{fig:ReImt2t3wt1}, one would observe a deformation of the
regions displayed in Fig.~\ref{fig:Arg:0MNS}. The analytical
interpretation can be now obtained from
Eqs.~(\ref{Eq:mt_1}-\ref{Eq:mt_3}), albeit in a less straightforward way.

\begin{figure}
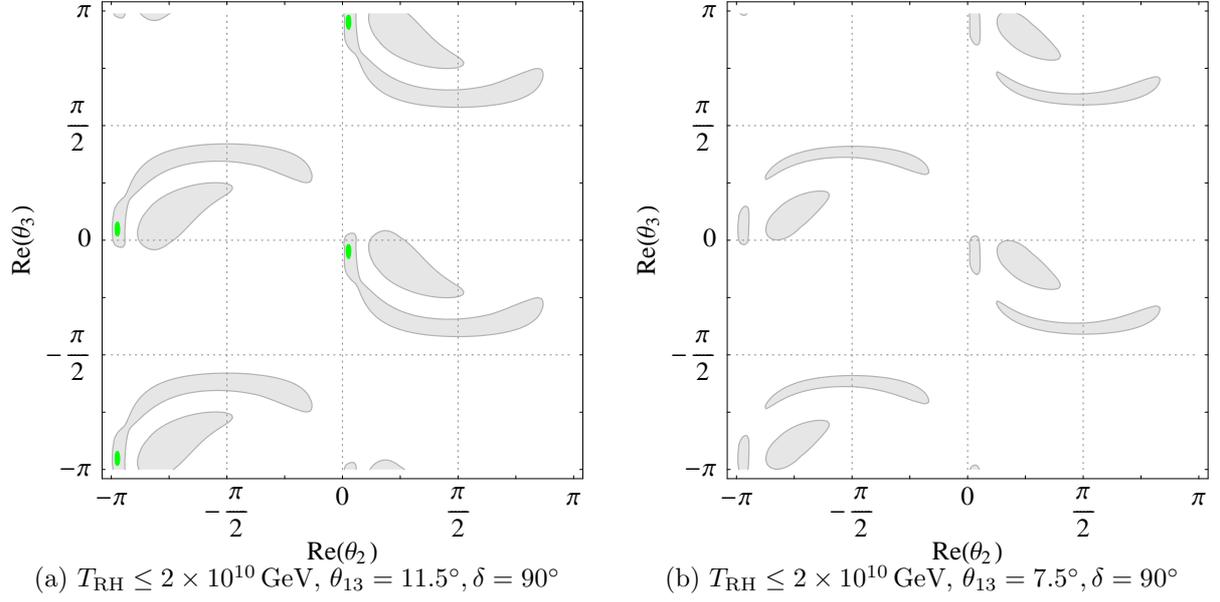

 \centering   \subfigure[ $T_{\mathrm{RH}} \le 2 \times
 10^{10}\, \text{GeV},\,\theta_{13}=11.5^\circ,\delta=90^\circ$]{$\CenterEps[0.895]{plot_t13_11k5_d_90_Re_t2_Re_t3_ScanN}$}  
 \;\;\;\, \subfigure[ $T_{\mathrm{RH}} \le 2 \times
 10^{10}\, \text{GeV},\,\theta_{13}=7.5^\circ,\delta=90^\circ$]{$\CenterEps[0.895]{plot_t13_7k5_d_90_Re_t2_Re_t3_ScanN}$}  
 \caption{\label{fig:t13_and_delta} 
Regions of $\mathrm{Re}(\theta_{2}) - \mathrm{Re}(\theta_{3})$
parameter space compatible with successful thermal leptogenesis 
with a bound $T_{\mathrm{RH}} \le 2 \times 10^{10}$ GeV.
In this example we have chosen real $R$-matrix angles and
$\theta_{1}=\varphi_1=\varphi_2=0^\circ$. 
Colour code as in Fig.~\ref{fig:ReImt2t3}.  
 }
\end{figure}

After having revisited leptogenesis scenarios where CP violation originated
solely from the complex $R$-matrix angles, let
us now consider the effects of having CP violation arising from the
$U_\text{MNS}$ phases.
This is especially appealing given
that, and contrary to the $R$-matrix angles, parameters like 
$\delta$ and $\theta_{13}$  
are likely to be observable in neutrino oscillation experiments.
Additionally, and as pointed out in 
Refs.~\cite{Pascoli:2006ie,Branco:2006ce}
(although in the context of the SM), these are examples of scenarios 
where there is a maximal connection between leptogenesis and 
low-energy CP violation.
We begin by addressing a scenario where the $R$-matrix angles are real
(but non-zero) and $\delta$ is the only source of CP violation.
For non-zero $\theta_{13}$ and $\delta$, 
Fig.~\ref{fig:t13_and_delta} illustrates the emergence of new regions
in the $\mathrm{Re}(\theta_{2}) - \mathrm{Re}(\theta_{3})$ 
parameter space, potentially
compatible with thermal leptogenesis in the presence of a bound
$T_{\mathrm{RH}} \le 2 \times 10^{10}$ GeV.
In this example we have chosen 
$\theta_{1}=\varphi_1=\varphi_2=0^\circ$. 
Notice that in the flavour-independent approximation,
leptogenesis would have been impossible for a real $R$-matrix.  
From Fig.~\ref{fig:t13_and_delta}, we find that for 
$\theta_{13}=11.5^\circ$ (the largest value experimentally allowed) 
and CP violating phase $\delta$ close to 
$\pi/2$ (which maximises the decay asymmetry $\varepsilon_{1,e}$),  
somewhat larger $\theta_2$ values could now be marginally allowed. 
In any case, the largest values of the BAU are still associated
with small $\theta_2$ and $\theta_3$. 
On the other hand, moving
away from the present upper bound on $\theta_{13}$, we find that the
scenario is even more compromised. 
In fact, for $\theta_{13} = 7.5^\circ$, only regions with BAU
differing from the WMAP range by a factor 5 survive. For smaller
$\theta_{13}$ values (namely $\theta_{13} \le 5^\circ$),
$n_\text{B}/n_\gamma$ is already well below $10^{-10}$.
Likewise, considering stricter bounds on $T_{\mathrm{RH}}$ would lead
to the disappearance of all the shaded regions of 
Fig.~\ref{fig:t13_and_delta}.

\begin{figure}
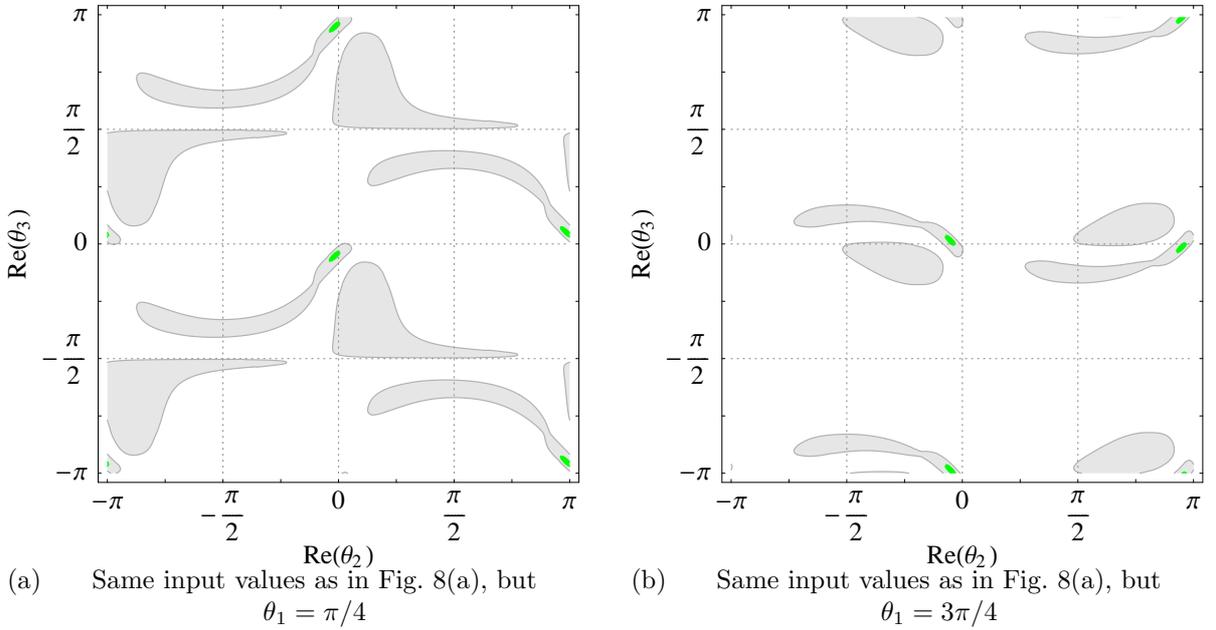

 \centering \subfigure[Same input values as in Fig.~\ref{fig:t13_and_delta}(a),
 but
 $\theta_1=\pi/4$]{$\CenterEps[0.895]{plot_t13_11k5_d_90_t1_Pio4_Re_t2_Re_t3_ScanN}$}    
 \;\;\;\, 
 \subfigure[Same input values as in Fig.~\ref{fig:t13_and_delta}(a),
 but
 $\theta_1=3\pi/4$]{$\CenterEps[0.895]{plot_t13_11k5_d_90_t1_Pi3o4_Re_t2_Re_t3_ScanN}$}    
 \caption{\label{fig:t13_and_delta_2} 
Example displayed in Figs.~\ref{fig:t13_and_delta}(a) and (c), but with
$\theta_1 \not =0^\circ$. Colour code as in Fig.~\ref{fig:ReImt2t3}. }
\end{figure}

In Fig.~\ref{fig:t13_and_delta_2}, we display the effect of non-zero $\theta_1$
on the allowed regions with non-zero $\theta_{13}$ and $\delta$.  
In the first example with $\theta_1 = \pi/4$ and $\theta_{13}=11.5^\circ$  
(Fig.~\ref{fig:t13_and_delta_2}(a)), we observe  
that in addition to a rotation of the 
allowed parameter space, the latter is somewhat enlarged. 

At this point, and from the examples so far considered, we are
led to the conclusion that, unless $\theta_{13}$ is found to be close to
its present upper bound, it is quite difficult to accommodate
viable BAU scenarios relying on $\delta$ as their only source of CP violation.
For stricter bounds on the reheat temperature, the latter
scenarios become increasingly more compromised.

\begin{figure}
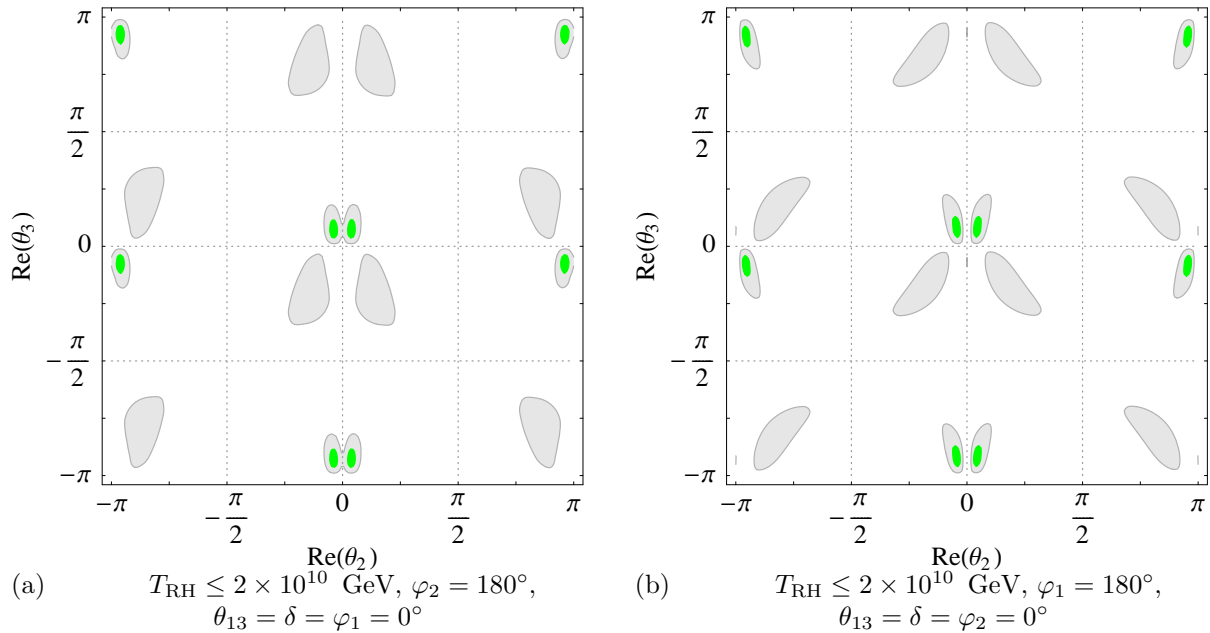

 \centering   
 \subfigure[$\quad \quad
 T_{\mathrm{RH}}\le 2\times 10^{10}\,~\text{GeV},\,\varphi_2=180^\circ,\,$
 $\quad \theta_{13}=\delta=\varphi_1=0^\circ$
 ]{$\CenterEps[0.895]{plot_t13_0_d_0_p2_180_Re_t2_Re_t3_ScanN}$} 
 \;\;\;\,  
 \subfigure[$\quad \quad
 T_{\mathrm{RH}}\le 2\times 10^{10}\,~\text{GeV},\,\varphi_1=180^\circ,\,$
 $\quad \theta_{13}=\delta=\varphi_2=0^\circ$
 ]{$\CenterEps[0.895]{plot_t13_0_d_0_p1_180_Re_t2_Re_t3_ScanN}$} 
 \caption{\label{fig:phi1phi2}
Regions of $\mathrm{Re}(\theta_{2})-\mathrm{Re}(\theta_{3})$
parameter space
compatible with successful thermal leptogenesis with 
$T_{\mathrm{RH}} \le 2 \times 10^{10}$ GeV.
In this example we have chosen real $R$, 
$\theta_{1}=0$ and $\theta_{13}=\delta=0^\circ$. 
Colour code as in Fig.~\ref{fig:ReImt2t3}. }
\end{figure}

In addition to $\delta$, there are other sources of CP violation
arising from the $U_\text{MNS}$ matrix, namely the Majorana phases 
$\varphi_1,\varphi_2$. In principle, the latter could also
provide the required CP violation for leptogenesis (see 
also Refs.~\cite{Blanchet:2006be,Pascoli:2006ie,Branco:2006ce}). 
For completeness, in Fig.~\ref{fig:phi1phi2}
we separately illustrate their role in generating a non-vanishing
BAU. To do so, we assume a real $R$-matrix,
$\delta=0^\circ$, and take $\varphi_{1(2)}=0^\circ$ on the left (right) panel. 
As seen from Fig.~\ref{fig:phi1phi2}, when CP violation
is exclusively arising from the Majorana phases
it is indeed possible to obtain marginally compatible BAU values.
Again, the most promising regions appear associated with small 
$\theta_2$ and $\theta_3$.
We also observe that $\varphi_1$ and $\varphi_2$ lead to somewhat
distinct regions on the
$\mathrm{Re}(\theta_{2})-\mathrm{Re}(\theta_{3})$ parameter
space. 
In this example we have again considered a more
relaxed bound for the reheat temperature, $T_{\mathrm{RH}} \le 2 \times
10^{10}$ GeV. As occurred for the cases investigated in
Fig.~\ref{fig:t13_and_delta}, stronger bounds on
$T_\text{RH}$ would imply that 
the generated BAU would also lie below
$10^{-10}$, so that the shaded regions of the  
$\mathrm{Re}(\theta_{2})-\mathrm{Re}(\theta_{3})$ parameter space
displayed in Fig.~\ref{fig:phi1phi2} would disappear.

\begin{figure}
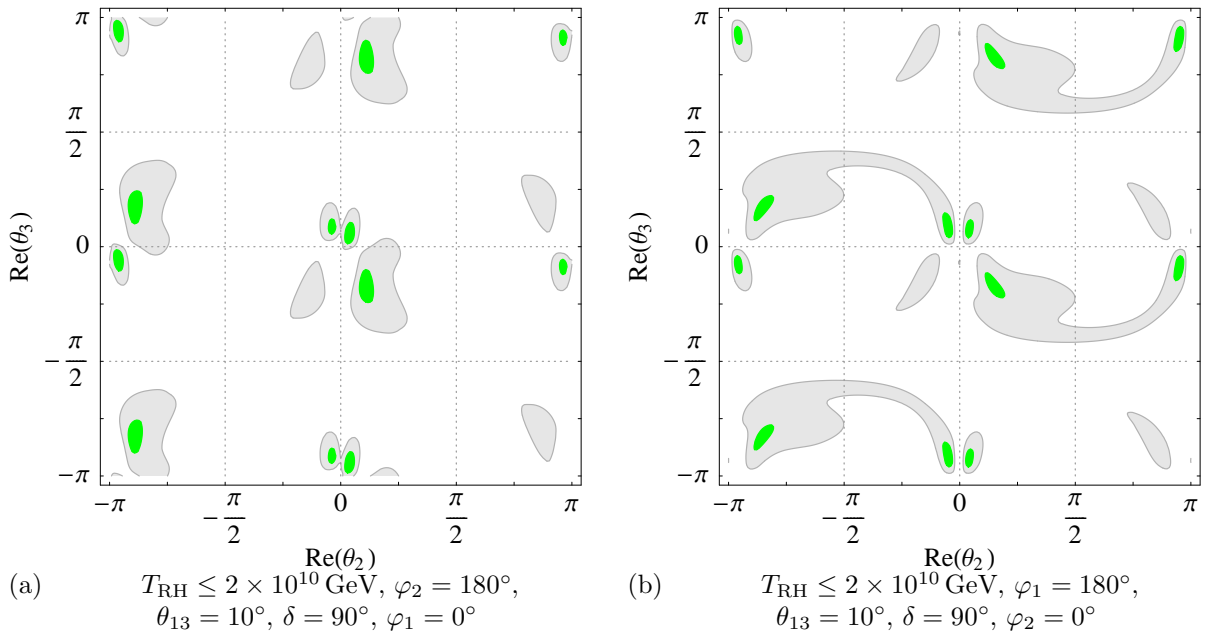

 \centering   
 \subfigure[$\quad 
T_{\mathrm{RH}} \le 2 \times 10^{10}\, \text{GeV},\,\varphi_2
 =180^\circ,$ $\quad \theta_{13}=10^\circ,\,\delta=90^\circ,\,\varphi_1
 =0^\circ$]{$\CenterEps[0.895]{plot_t13_10_d_90_p2_180_Re_t2_Re_t3_ScanN}$}
\;\;\;\;\subfigure[$\quad 
T_{\mathrm{RH}} \le 2 \times 10^{10}\, \text{GeV},\,\varphi_1
 =180^\circ,$ $\quad \theta_{13}=10^\circ,\,\delta=90^\circ,\,\varphi_2
 =0^\circ$]{$\CenterEps[0.895]{plot_t13_10_d_90_p1_180_Re_t2_Re_t3_ScanN}$}
\\ 
\caption{\label{fig:phi1phi2_2}
Same examples as in Fig.~\ref{fig:phi1phi2}, but with
$\theta_{13}=10^\circ$ and $\delta=90^\circ$.} 
\end{figure}

Albeit it is pedagogical to consider the individual role of each 
$U_\text{MNS}$ phase regarding BAU, in the most
general case $\delta$, $\varphi_1$ and $\varphi_2$ can be
simultaneously non-vanishing. In fact, and as shown in  
Fig.~\ref{fig:phi1phi2_2}, the Majorana phases can slightly
improve the BAU allowed regions associated with $\delta$
(and $\theta_{13}$). We recall that, as discussed in relation 
with Fig.~\ref{fig:t13_and_delta},
values $\theta_{13} \leq 5^\circ$ failed to induce 
$n_\text{B}/n_\gamma > 10^{-10}$.
Comparing Fig.~\ref{fig:phi1phi2_2} with
Fig.~\ref{fig:t13_and_delta}(c), we observe that for the choice
of $\varphi_{1,2} = 180^\circ$ the regions where $n_\text{B}/n_\gamma
\in [3 \times 10^{-10},\,5.9 \times 10^{-10}]$ have become larger.
Even though the WMAP range cannot be accounted for, it is nevertheless
clear that the joint effect of the different $U_\text{MNS}$ phases
translates in an improved scenario.

Even though flavour-dependent thermal leptogenesis opens the
possibility to generate the observed BAU exclusively from the
$U_\text{MNS}$ CP violating phases, this may not be the most general
nor the most successful scenario. The analysis of the present section
lends a strong support to the latter statement. As we have found, it
is quite difficult to encounter viable BAU scenarios associated with
only low-energy CP violation. Moreover, if the given SUSY model 
implies a more stringent bound 
on the reheat temperature, BAU solely from 
$U_\text{MNS}$ phases becomes almost inviable.

Recall that in the most general case, the $R$-matrix angles are also
complex and that, as seen from
Figs.~\ref{fig:ReImt2t3}, \ref{fig:ReImt2t3wt1} 
and \ref{fig:Arg:0MNS}, there are important regions in the 
$\theta_2$-$\theta_3$ parameter space where one can easily have
compatibility with the WMAP range. In addition, $R$-matrix phases allow
for viable BAU even under a bound $T_{\mathrm{RH}} \le 5 \times 10^{9}$ GeV.
Thus, it is important to investigate the simultaneous effect of all CP
violating phases. In particular, one wonders to which extent the
$U_\text{MNS}$ phases can affect the BAU predictions from the
$R$-matrix phases, and vice-versa.

\begin{figure}
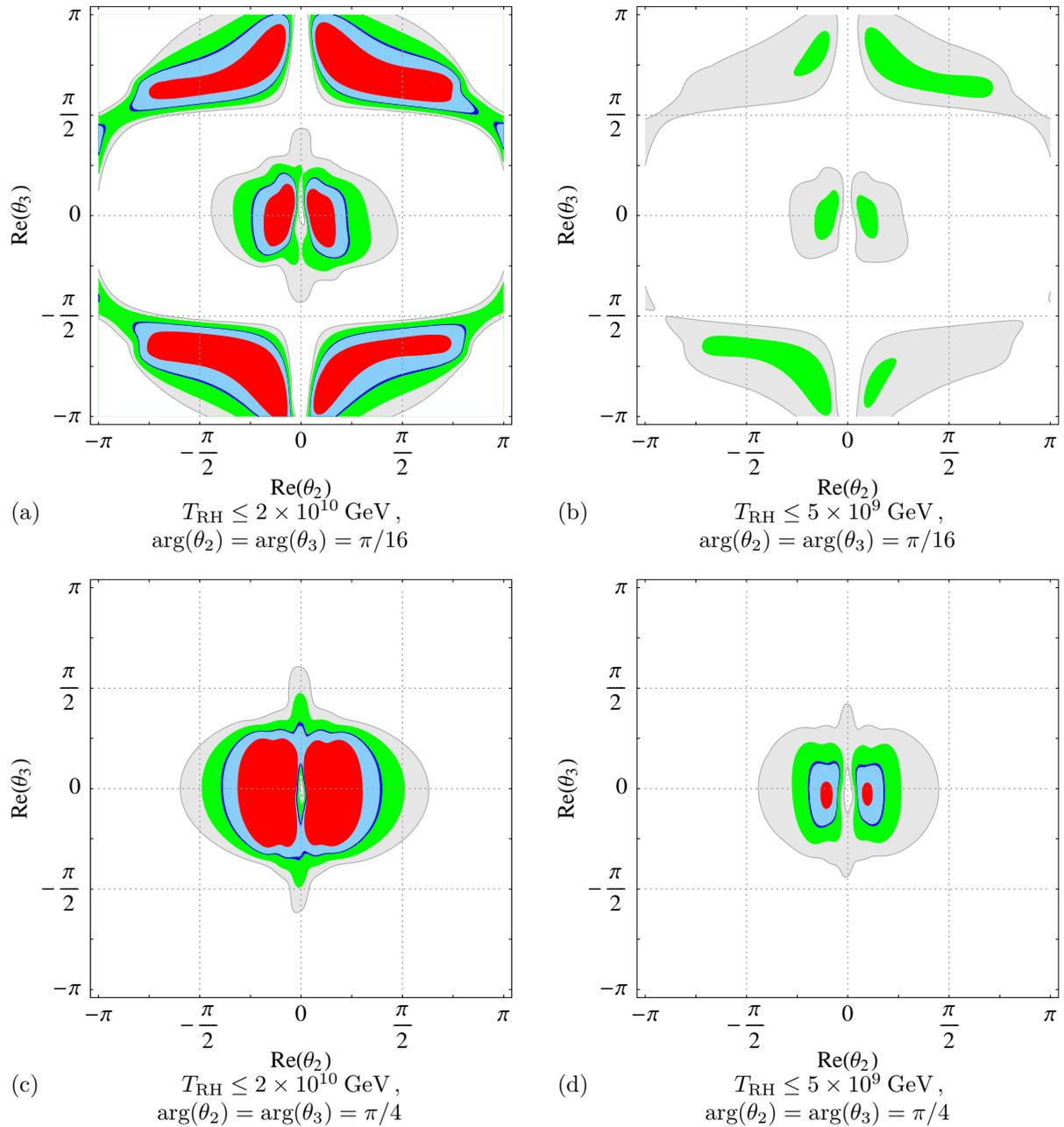

 \centering
 \subfigure[$ \quad  T_{\mathrm{RH}} \le 2 \times
 10^{10}\:\mbox{GeV}\,,$ $\,\mbox{arg}(\theta_2) = \mbox{arg}(\theta_3) =
 \pi/16$
 ]{$\CenterEps[0.895]
{plot_t13_10_d_90_p1_180_Argt2_Pio16_Argt3_Pio16_Re_t2_Re_t3_ScanN}$}
 \;\;\;\,
 \subfigure[$ \quad T_{\mathrm{RH}} \le 5 \times
 10^{9}\:\mbox{GeV}\,,$ $\,\mbox{arg}(\theta_2) = \mbox{arg}(\theta_3) =
 \pi/16$
 ]{$\CenterEps[0.895]
{plot_t13_10_d_90_p1_180_Argt2_Pio16_Argt3_Pio16_Re_t2_Re_t3_ScanN_5_9}$}\\
 \subfigure[$ \quad T_{\mathrm{RH}} \le 2 \times
 10^{10}\:\mbox{GeV}\,,$ $\,\mbox{arg}(\theta_2) = \mbox{arg}(\theta_3) =
 \pi/4$
 ]{$\CenterEps[0.895]
{plot_t13_10_d_90_p1_180_Argt2_Pio4_Argt3_Pio4_Re_t2_Re_t3_ScanN}$}
 \;\;\;\,
 \subfigure[$ \quad T_{\mathrm{RH}} \le 5 \times
 10^{9}\:\mbox{GeV}\,,$ $\,\mbox{arg}(\theta_2) = \mbox{arg}(\theta_3) =
 \pi/4$
 ]{$\CenterEps[0.895]
{plot_t13_10_d_90_p1_180_Argt2_Pio4_Argt3_Pio4_Re_t2_Re_t3_ScanN_5_9}$}
 \caption{\label{fig:Arg}
Regions of $\mathrm{Re}(\theta_{2})-\mathrm{Re}(\theta_{3})$
parameter space
compatible with successful thermal leptogenesis. We take
$\theta_{13}=10^\circ,\,\delta=90^\circ,$ and $\varphi_1 =180^\circ$. We
consider $\arg (\theta_2) = \arg (\theta_3) = \pi/16$ for panels (a) and (b),
and $\arg (\theta_2) = \arg (\theta_3) = \pi/4$ for panels (c) and (d).
$T_{\mathrm{RH}} \le 2 \times 10^{10}$ GeV ((a) and (c))
while $T_{\mathrm{RH}} \le 5 \times 10^{9}$ GeV ((b) and (d)).
Colour code as in Fig.~\ref{fig:ReImt2t3}.
 }
\end{figure}

In Fig.~\ref{fig:Arg} we display the outcome of taking, in
addition to the CP sources considered in Fig.~\ref{fig:phi1phi2_2},
non-vanishing values for the arguments of $\theta_2$ and
$\theta_3$, namely $\pi/16$ (upper) and $\pi/4$ (lower). 
The results are again depicted in the
$\mathrm{Re}(\theta_{2})-\mathrm{Re}(\theta_{3})$ plane, and we
assume two bounds for the reheat temperature,  
$T_{\mathrm{RH}} \le 2 \times 10^{10}$ GeV (left) and
$T_{\mathrm{RH}} \le 5 \times 10^{9}$ GeV (right).
It is manifest from the comparison of 
Fig.~\ref{fig:Arg:0MNS} with Fig.~\ref{fig:Arg} 
that the predictions for the 
$\mathrm{Re}(\theta_{2})-\mathrm{Re}(\theta_{3})$ plane are hardly
affected by considering non-vanishing values for the
$U_{\mathrm{MNS}}$ phases. In the case of $\mbox{arg} (\theta_{2,3})=\pi/16$,
and even for nearly maximal values of
$\theta_{13}$, one only observes a small distortion of the regions
associated with small $|s_2|$ and $|s_3|$, 
and a deformation of the regions associated with large 
$|s_2|$ and $|s_3|$ (c.f.~Fig.~\ref{fig:Arg}(b)).
For larger arguments of $\theta_2$ and $\theta_3$ ($\pi/4$),
comparing Fig.~\ref{fig:Arg}(d) with Fig.~\ref{fig:ReImt2t3wt1}(a) 
implies that $\theta_{13}$ and $\delta$ have had
virtually no effect on the shape of the BAU compatible region, 
which is determined by the right-handed neutrino
complex parameters $\theta_2$ and $\theta_3$.
Finally, notice that for $T_{\mathrm{RH}} \le 5 \times
 10^{9}$ GeV, it is possible to nearly reach the WMAP range for 
$\mbox{arg} (\theta_{2,3})$ as small as $\pi/16$.  Assuming larger values for
the arguments of the $R$-matrix angles allows to encounter wider
regions where one still has compatibility with WMAP observations (panel (d)).
However, this again favours
the region of small $|s_2|$ and $|s_3|$.

In a sense, Fig.~\ref{fig:Arg} provides an illustrative summary of our
analysis. Firstly, it confirms the supposition that flavour effects in
the Boltzmann equations indeed lead to the occurrence of new regions
where BAU is viable (due to the already mentioned reduced
washout). In addition, from its comparison to
Fig.~\ref{fig:ReImt2t3wt1}, it is also manifest that, in general, the
leading role in BAU appears to be played 
by the $R$-matrix complex angles, and
not by the $U_{\mathrm{MNS}}$ phases (which in turn may compromise the
possible bridges that could otherwise be established between
low-energy CP violation and BAU).
Moreover, Fig.~\ref{fig:Arg} is a clear example of the impact of the reheat
temperature in severely constraining the SUSY seesaw parameters. The
stricter the bounds on $T_{\mathrm{RH}}$, the more favoured are
regions with small $\text{Re}(\theta_{2,3})$ and large $\mbox{arg}
(\theta_{2,3})$. 

\section{Summary and conclusions}\label{concl}

In this study we have investigated the constraints on the SUSY
seesaw parameter space arising from flavour-dependent
thermal leptogenesis in the MSSM in the presence of upper bounds on the reheat
temperature of the early Universe. 
In the temperature range here considered, both 
tau- and muon-flavours are in thermal equilibrium, so that the full 
flavour-dependence was taken into account.
In order to calculate the efficiency factor for thermal
leptogenesis, we have extended the flavour-dependent Boltzmann
equations~\cite{davidsonetal,nardietal,Abada:2006ea},
which were adapted to the MSSM case in~\cite{Antusch:2006cw},
to include reheating (following Ref.~\cite{Giudice:2003jh}).
Parameterising the solutions to the seesaw equation by means of a
complex orthogonal matrix $R$~\cite{Casas:2001sr}, we have analysed
which regions of the seesaw parameter space generically enable optimal
efficiency and/or optimal decay asymmetries for leptogenesis.

We have discussed several differences between the flavour-independent
approximation and the correct flavour-dependent treatment of thermal
leptogenesis. These are extensive, and, together with the bounds from
the reheat temperature, lead to interesting new constraints on the SUSY
seesaw parameter space.

Considerations on $T_\text{RH}$ give rise to the first constraints, 
in the sense that a dramatic drop in the
efficiency takes place for $m_{N_1} \gg T_\text{RH}$ (as much as three
orders of magnitude for $m_{N_1} \approx 10 \,T_\text{RH}$).
Since we have assumed that only the decays of the lightest
right-handed neutrino were relevant for the lepton asymmetry, no
bounds on the masses $m_{N_2}$ and $m_{N_3}$ were derived.
Assuming an optimal regime for the decay asymmetry and for the
efficiency, as well as that the inflaton only decays into MSSM particles 
(and not directly into right-handed (s)neutrinos), 
the requirement of a successful BAU leads to lower bounds
on $m_{N_1}$ as well as on the reheat temperature. In particular, we
have found $m_{N_1}^\text{min} \approx 1.5 \times 10^9$ GeV and
$T_\text{RH}^\text{min} \approx 1.9\times 10^9$ GeV,
similar to the results obtained in the flavour-independent 
approximation~\cite{Giudice:2003jh}.
On the other hand, in the presence of upper
bounds on the reheat temperature (from dark matter relic abundance
considerations), an upper bound on $m_{N_1}$ can also be inferred.
In order to illustrate the impact of reheating,
we have considered two examples for
$T_\text{RH}$, corresponding to mildly and strongly constrained
scenarios: $T_\text{RH} \lesssim 2 \times 10^{10}$ GeV, and
$T_\text{RH} \lesssim 5 \times 10^{9}$ GeV.
Regarding the upper bound on $m_{N_1}$, the latter bounds respectively yield
$m^\text{max}_{N_1} \approx 1.4 \times 10^{11}$ GeV, and 
$m^\text{max}_{N_1} \approx 1.9 \times 10^{10}$ GeV.
This leads to viability windows for the mass of the lightest right-handed
neutrino.

Regarding the light neutrino masses, namely $m_{\nu_1}$,
in general there is no upper bound from flavour-dependent thermal
leptogenesis.
Nevertheless, increasing $m_{\nu_1}$ within the present allowed
experimental upper bounds (of roughly $0.5$ eV), generically results in
a reduced BAU. 
Furthermore, in the presence of strong $T_\text{RH}$ bounds, quasi-degenerate
light neutrino masses (via the type-I seesaw mechanism) become disfavoured.

As in the flavour-independent approximation,
considerations on the washout parameters generically favour 
the region of small $|\sin \theta_2|$ and $|\sin \theta_3|$.
However, and in clear contrast to the flavour-independent approximation, 
new regions with optimal flavour-dependent washout parameters have
emerged, in association with large values of $|\sin \theta_2|$ and
$|\sin \theta_3|$. In any case, 
in order to produce sufficient BAU in the presence 
of mild (or even strong) constraints on the reheat temperature,
the decay asymmetries must also be close
to their optimal values.
Regarding the flavour-dependent decay asymmetries, in the general case
of complex $R$ matrix (but considering the limit of a real
$U_\text{MNS}$), we generically recovered the main results of the
one-flavour computation (in the
sense that the favoured regions still corresponded to 
small $|\sin \theta_2|$ and $|\sin \theta_3|$, albeit slightly enlarged). 
Another important result of our analysis concerns the effects of the
reheat temperature, which were clearly manifest. In fact, taking
stronger bounds on $T_\text{RH}$ leads to a significant reduction in
the BAU allowed regions of the
$\mathrm{Re} (\theta_2)$-$\mathrm{Im} (\theta_2)$ and
$\mathrm{Re} (\theta_3)$-$\mathrm{Im} (\theta_3)$ parameter spaces
(even to the disappearance of the WMAP compatible regions).
In particular, for $T_\text{RH} \lesssim 5 \times 10^{9}$ GeV,
we have seen that $\theta_3$ cannot exclusively account 
for the observed WMAP results.

In flavour-dependent leptogenesis, a potentially important role can also be
played by the $U_\text{MNS}$ phases. 
In principle, viable BAU scenarios could be obtained in the presence of a
CP-conserving $R$-matrix, with the required amount of CP violation
stemming either from the Dirac phase $\delta$, or from the Majorana
phases $\varphi_1$ and $\varphi_2$.
In the SM, this situation has been discussed in
Refs.~\cite{Pascoli:2006ie,Branco:2006ce}. 
However, the constraints on the seesaw parameters in the 
MSSM are expected to differ from the SM case, 
since for the temperatures (and values of $\tan \beta$)
under consideration, all flavours are separately treated in the MSSM
Boltzmann equations, whereas only the {\small$\tau$}-flavour is separately
considered in the SM case.  

Exclusively relying on the phase $\delta$ and on $\theta_{13}$ (under
the standard parameterisation of the $U_\text{MNS}$ matrix) is a
phenomenologically challenging choice, since these are the most likely
(yet) unknown $U_\text{MNS}$ parameters to be experimentally measured.
However, we have verified 
that even with $\delta=\pi/2$, for values of $\theta_{13} <
11.5^\circ$ (the present experimental limit) the obtained values of
the baryon asymmetry are only marginally compatible with observation
(when large theoretical uncertainties are allowed for).
By themselves, and even in the limit $\theta_{13}=0^\circ$,
both Majorana phases, $\varphi_1$ and $\varphi_2$, could in principle account 
for BAU. However, and similar to $\delta$, only marginal consistency with 
observations can be reached. In both cases
the impact of the reheating temperature becomes manifest, since lower
values of $T_\text{RH}$ can render these scenarios inviable.
We also note that in the presence of small $\mbox{arg}(\theta_i)$, the
BAU generated from CP violation in the right-handed sector dominates
over the contributions from low-energy phases. Thus, 
the sensitivity to the $U_\text{MNS}$ CP violating phases is lost. 
In the limit of very strict bounds on the reheat temperature, 
one is thus compelled to take into account complex $R$ as 
an additional source of CP violation. 

In summary, we have investigated which regions of 
the SUSY seesaw parameter space 
are favoured by flavour-dependent thermal leptogenesis, when bounds on  
the reheat temperature are taken into account. 
For mildly constrained $T_\text{RH}$ (e.g.\ $T_\text{RH} \lesssim 2 \times
10^{10}$ GeV), compatibility with the BAU observed by WMAP can be obtained for
extensive new regions of the $\theta_2$-$\theta_3$ parameter space, which were
previously disfavoured in the flavour-independent approximation.   
On the other hand, focusing on (normal) hierarchical light and heavy neutrinos, the scenario where only CP violation 
from the $U_\text{MNS}$ is
considered (real $R$), turns out to be only marginally consistent, even for
$\theta_{13}=10^\circ$, and under mild bounds on $T_{\mathrm{RH}}$.
Stricter $T_\text{RH}$ bounds (e.g.\ $T_\text{RH} \lesssim
5 \times 10^{9}$ GeV) strongly motivate that CP is (also) violated 
in the right-handed neutrino sector.
While extensive regions of the $\theta_2$-$\theta_3$ parameter space can
produce BAU close to the WMAP range in this case, the favoured seesaw parameter
space, clearly consistent with observations, 
is that of small values of $|\sin \theta_2|$ and $|\sin \theta_3|$.

Given the attractiveness of the mechanism of thermal leptogenesis, and the
interesting constraints it can provide, it would be desirable to further
refine the computation of the baryon asymmetry.
Together with the expected improved bounds from LFV, electric 
dipole moments and other related observables, leptogenesis may offer 
valuable information on the right-handed neutrino masses and mixings.  

\section*{Acknowledgements}
We are grateful to F.~R.~Joaquim, S.~F.~King and A.~Riotto
for enlightening discussions.
We also thank E.~Arganda and M.~J.~Herrero for several important remarks.
The work of S.~Antusch was supported by the EU 6$^\text{th}$
Framework Program MRTN-CT-2004-503369 ``The Quest for Unification:
Theory Confronts Experiment''. 
The work of A.~M.~Teixeira has been supported by the French ANR
project PHYS@COL\&COS and by HEPHACOS
`` Fenomenolog\'{\i}a de las Interacciones Fundamentales: Campos,
Cuerdas y Cosmolog\'{\i}a'' P-ESP-00346. 
A.~M.~Teixeira further acknowledges the support of ``Acci\'on Integrada
Hispano-Francesa''.

\section*{Appendix}
\appendix

\renewcommand{\thesection}{\Alph{section}}
\renewcommand{\thesubsection}{\Alph{section}.\arabic{subsection}}
\def\theequation{\Alph{section}.\arabic{equation}}
\renewcommand{\thetable}{\arabic{table}}
\renewcommand{\thefigure}{\arabic{figure}}
\setcounter{section}{0}
\setcounter{equation}{0}

\section{Boltzmann equations with reheating}\label{App:BoltzmannReheating}

The efficiency factor for thermal leptogenesis introduced in
Section~\ref{Sec:FML} is calculated from the flavour-dependent
Boltzmann equations~\cite{barbieri,davidsonetal,nardietal,Abada:2006ea},
generalised to the MSSM case~\cite{Antusch:2006cw}. Regarding
reheating, one follows the simple, yet convenient approach of
Ref.~\cite{Giudice:2003jh}, where the effects of reheating are
described by a single parameter, $T_\text{RH}$.
The limitations of the several
approaches were summarised in Section~\ref{Sec:limitations}.
In this appendix, aiming at completeness,
we present some technical details.

The Boltzmann equations, with $z = m_{N_1}/T$, can be written as
\cite{Antusch:2006cw,Giudice:2003jh}:
\begin{eqnarray}
\label{0s}
 Z \,\frac{\mathrm{d} \rho_{\phi}}{\mathrm{d} z} &=&
- \frac{3}{z} \,\rho_{\phi} - \frac{\Gamma_\phi}{H\, z} \,\rho_{\phi}
\; ,\\
\label{1s} Z \,X\,\frac{\mathrm{d} Y_{N_1}}{\mathrm{d} z} &=&
\frac{3 \,(Z - 1)\, X}{z} \, Y_{N_1}\, -\,  2 K\,z \,\frac{K_1
  (z)}{K_2 (z)}\, f_1 (z) \,(Y_{N_1} - Y^\mathrm{eq}_{N_1}) \; ,\\
\label{2s}  Z \,X\,\frac{\mathrm{d} Y_{\widetilde N_1}}{\mathrm{d} z} &=&
\frac{3 \,(Z - 1)\, X}{z} \, Y_{\widetilde N_1}\,-\,  2 K\,z
\,\frac{K_1 (z)}{K_2 (z)}\, f_1 (z) \,(Y_{\widetilde N_1} -
Y^\mathrm{eq}_{\widetilde N_1}) \; ,\\
\label{3s}  Z \,X\,\frac{\mathrm{d} \hat Y_{\Delta_\alpha}}{\mathrm{d} z} &=&
- \; 2\,\varepsilon_{1,\alpha} \,K\, z \,
\frac{K_1 (z)}{K_2 (z)}\,  f_1 (z) \,\left[ (Y_{N_1} -
  Y^\mathrm{eq}_{N_1}) \,+\,(Y_{\widetilde N_1} -
  Y^\mathrm{eq}_{\widetilde N_1})\right]
 \\
&&+\;
\frac{3 \,(Z - 1)\, X}{z} \,Y_{\alpha}
\;+\;
K_{\alpha} \, z\, \frac{K_1 (z)}{K_2 (z)} \, f_2 (z) \,
 \frac{\sum_\beta A_{\alpha\beta}\,\hat Y_{\Delta_\beta}}
{Y^\mathrm{eq}_{\alpha}}
 \,(Y^\mathrm{eq}_{N_1}+Y^\mathrm{eq}_{\widetilde N_1})
\; \nonumber .
\end{eqnarray}
The above equations should be solved from $z = m_{N_1}/T^\mathrm{max}$
to ``infinity'' (i.e. \ $z \gg 1$).

Let us now address each of the quantities appearing in
Eqs.~(\ref{0s} - \ref{3s}).
First of all, let us comment on the effects of reheating.
$\rho_{\mathrm{R}}$ and $\rho_{\phi}$ are the radiation energy density
and the energy density from the coherent oscillations of the reheating
scalar field $\phi$. The reheating temperature is given by
\begin{eqnarray}\label{Eq:TRH_fromGamma_phi}
T_{\mathrm{RH}} = \left( \frac{45\, \Gamma^2_\phi \,
  M^2_\mathrm{Pl}}{4 \pi^3 g_*} \right)^{\frac{1}{4}} ,
\end{eqnarray}
where $\Gamma_\phi$ is the decay rate of $\phi$,
$M_\mathrm{Pl}$ is the Planck scale, and $g_*=228.75$ was already
introduced in Eq.~(\ref{eq:Yeq}).
During reheating, $\rho_{\phi}$ dominates over $\rho_{\mathrm{R}}$.
In the Boltzmann equations, reheating is taken into account by means of
\begin{eqnarray}
Z = 1 - \frac{\Gamma_\phi \,\rho_\phi}{4 H \rho_\mathrm{R}} \;,
\end{eqnarray}
which is equal to $0$ when the maximal reheat temperature
$T^\mathrm{max}$ is reached (corresponding to our initial
conditions), and which becomes $1$ after reheating.
In the limit $Z \to 1$ (and $X \to 1$ - see definition below, in
Eq.~(\ref{Eq:Xdef})), we recover the MSSM equations
without reheating, as given in~\cite{Antusch:2006cw}.
At the maximal temperature $T^\mathrm{max}$, the energy density
$\rho_\phi$ can be calculated from the condition $Z=0$, using
$\rho_\mathrm{R} = (m_{N_1}/z)^4 \,\pi^2 g_* /30$ together with 
Eq.~(\ref{Eq:HwithPhi}).
Notice that Eq.~(\ref{Eq:TRH_fromGamma_phi}) allows to extract
$\Gamma_\phi$ (appearing in Eq.~(\ref{0s}))
as a function of the reheat temperature,
\begin{eqnarray}
\Gamma_\phi = \left( \frac{T_\mathrm{RH}^4}{M_\mathrm{Pl}^2} \frac{4
  \pi^3 g^*}{45} \right)^{\frac{1}{2}}.
\end{eqnarray}
We would like to stress that in specific models of reheating after
inflation, the prospects for leptogenesis could be significantly
different. Nevertheless, this set of Boltzmann equations
(Eqs.~(\ref{0s} - \ref{3s})) simulates the
generic constraints arising for thermal leptogenesis from bounds on
the reheat temperature for a large class of scenarios.

$\hat Y_{\Delta_\alpha}$ are defined as
$\hat Y_{\Delta_\alpha}=Y_B/3 - Y_{L_\alpha}$,
with $Y_{L_\alpha}$ being the total (particle and sparticle) lepton number
densities for a flavour {\small$\alpha$}. 
$Y_{\alpha,\widetilde \alpha}$ are the
densities of the (s)lepton doublets and $Y_{N_1,\widetilde N_1}$
are the densities of the right-handed (s)neutrinos.
The corresponding equilibrium number densities (in the Boltzmann
approximation) are given by
\begin{eqnarray}
 Y^{\mathrm{eq}}_{\ell}  \!\!\!&\approx& \!\!\! Y^{\mathrm{eq}}_{\widetilde
\ell} \approx
 \frac{45 \,\zeta (3)}{ \pi^4
 g_* }\frac{3}{4} \; , \quad \quad
 Y^{\mathrm{eq}}_{N_1}(z) \approx Y^{\mathrm{eq}}_{\widetilde N_1}(z) \approx
 \frac{45 \,\zeta (3)}{ 2 \pi^4
 g_* }\frac{3}{4} \,z^2\, K_2 (z)
 \;,
\end{eqnarray}
with $K_2 (z)$ (and $K_1 (z)$) being 
the modified Bessel functions of the second kind.

The matrix $A$, which appears in the washout term in Eq.~(\ref{3s}), is
defined via the relation
$\hat Y_{\alpha} = \sum_\beta A_{\alpha\beta} \, \hat Y_{\Delta_\alpha}$, with
$\hat Y_\alpha \equiv Y_\alpha + Y_{\widetilde \alpha}$ being the combined
densities for lepton and slepton doublets.
Below $(1+\tan^2 \beta)\times 10^9$ GeV, where the Boltzmann equations are
solved for the individual asymmetries $\hat Y_{\Delta_e}$, $\hat
Y_{\Delta_\mu}$ and $\hat Y_{\Delta_\tau}$, $A$ is given
by~\cite{Antusch:2006cw}
\begin{eqnarray}
A^\mathrm{MSSM} =
\begin{pmatrix}
-93/110 & 6/55 & 6/55 \\
3/40 & -19/30 & 1/30 \\
3/40 & 1/30 & -19/30
\end{pmatrix}.
\end{eqnarray}

The final lepton asymmetry in each flavour is governed by
$T_\mathrm{RH}$ and by three sets of parameters:
$\varepsilon_{1,\alpha},K_{\alpha}$ and $K$. 
The parameters
$\varepsilon_{1,\alpha}$, $\varepsilon_{1,\widetilde \alpha}$,
$\varepsilon_{\widetilde 1,\alpha}$ and $\varepsilon_{\widetilde
  1,\widetilde \alpha}$ denote the asymmetries for the decays of
neutrino into Higgs and lepton, neutrino into Higgsino and slepton,
sneutrino into Higgsino and lepton, and sneutrino into Higgs and
slepton, respectively. They are defined as
\begin{eqnarray}\label{Eq:EpsMSSM_def}
\varepsilon_{1,\alpha} \!\!\!&=&\!\!\!
\frac{
\Gamma_{N_1 \ell_\alpha} - \Gamma_{N_1 \overline \ell_\alpha}
}{
\sum_\alpha (\Gamma_{N_1 \ell_\alpha} + \Gamma_{N_1 \overline \ell_\alpha})
}\; , \quad
\varepsilon_{1,\widetilde \alpha} =
\frac{
\Gamma_{N_1 \widetilde \ell_\alpha} - \Gamma_{N_1
\widetilde{\ell}_\alpha^*}
}{
\sum_\alpha (\Gamma_{N_1 \widetilde \ell_\alpha} + \Gamma_{N_1
\widetilde{\ell}_\alpha^*})
}\; , \nonumber  \\
\varepsilon_{\widetilde 1,\alpha} \!\!\!&=&\!\!\!
\frac{
\Gamma_{\widetilde N^*_1 \ell_\alpha} - \Gamma_{\widetilde N_1 \overline
  \ell_\alpha}
}{
\sum_\alpha (\Gamma_{\widetilde N^*_1 \ell_\alpha} +
\Gamma_{\widetilde N_1 \overline \ell_\alpha})
}\; , \quad
\varepsilon_{\widetilde 1,\widetilde \alpha} =
\frac{
\Gamma_{\widetilde N_1 \widetilde \ell_\alpha} - \Gamma_{\widetilde N^*_1
\widetilde{\ell}_\alpha^*}
}{
\sum_\alpha (\Gamma_{\widetilde N_1 \widetilde \ell_\alpha} +
\Gamma_{\widetilde N^*_1
\widetilde{\ell}_\alpha^*})
}\;,
\end{eqnarray}
with $\Gamma$ the decay rates of (s)neutrinos with (s)leptons in the final
states. In the MSSM, the four decay asymmetries are equal,
$\varepsilon_{1,\alpha} =
\varepsilon_{1,\widetilde \alpha} =
\varepsilon_{\widetilde 1,\alpha} =
\varepsilon_{\widetilde 1,\widetilde \alpha}$,
and given by Eq.~(\ref{Eq:EpsMSSM}).
The parameters $K_{\alpha}$ control the washout processes for the
asymmetry in an individual lepton flavour {\small $\alpha$}, and $K$ controls
the source of right-handed neutrinos in the thermal bath.
In analogy to the case without reheating, they are given by
\begin{eqnarray}\label{Eq:Kaa}
K_{\alpha} \equiv \frac{\Gamma_{N_1 \ell_\alpha} +\Gamma_{N_1
    \overline\ell_\alpha}}{H_0(m_{N_1})}\;,\quad
K \equiv \sum_\alpha  K_\alpha\;,\quad
K_\alpha = K
\frac{(\lambda_{\nu}^{\dagger})_{1\alpha}(\lambda_{\nu})_{\alpha 1}}
{(\lambda_{\nu}^{\dagger}\lambda_{\nu})_{11}}\;,
\end{eqnarray}
where $H_0(m_{N_1})$ is the ``fictitious'' 
Hubble parameter (without reheating).
The latter is computed without taking into account
$\rho_\phi$ at $T=m_{N_1}$, and is given by
$H_0(m_{N_1}) \approx 1.66 \sqrt{g_*} m_{N_1}^2/M_\mathrm{Pl}$.
In the presence of $\rho_\phi$, the Hubble parameter is modified to
\begin{eqnarray}\label{Eq:HwithPhi}
H = \left[\frac{8 \pi}{3}  (\rho_{\mathrm{R}} + \rho_{\phi})
  \right]^{\frac{1}{2}} \frac{1}{M_\mathrm{Pl}}\;,
\end{eqnarray}
and in order to match the $H$ (real) and $H_0$ (``fictitious'') 
Hubble parameters, one
introduces the quantity $X$,
\begin{eqnarray}\label{Eq:Xdef}
X \equiv \left(\frac{{\rho_{\mathrm{R}} +
    \rho_{\phi}}}{{\rho_{\mathrm{R}}}}\right)^{\frac{1}{2}} \,.
\end{eqnarray}
We further notice that the parameters $K_{\alpha}$ are
related to $\widetilde m_{1,\alpha}$, introduced in Eq.~(\ref{Eq:mtildeaa}), as
\begin{eqnarray}\label{Eq:Ka_mt}
\widetilde m_{1,\alpha} = K_{\alpha} \, m^* , \; \mbox{with}\;
m^*\approx \sin^2(\beta) \times 1.58 \times 10^{-3} \ {\rm eV} \,\, 
\cite{Antusch:2006cw}\,.
\end{eqnarray}

Finally, the  function $f_1(z)$ accounts for the
presence of $\Delta L=1$ scatterings
and $f_2(z)$ accounts for scatterings in the washout
term of the asymmetry.
They are defined as
\begin{eqnarray}
\gamma_D + \gamma_{\mathrm{S},\Delta L = 1} \equiv
\gamma_D f_1 \; , \quad
\frac{\gamma^{\alpha}_D}{2}  + \gamma^{\alpha}_{\mathrm{W},\Delta L = 1}
\equiv
\gamma^{\alpha}_D f_2\; ,
\end{eqnarray}
where $\gamma_D$ is the thermally averaged total decay rate of $N_1$ and
$\gamma_{\mathrm{S},\Delta L = 1}$ represents the rates for the
$\Delta L = 1$ scattering.  The corresponding flavour-dependent rates
for washout processes involving the lepton flavour $\alpha$ are
denoted by $\gamma^{\alpha}_D$ (from inverse decays involving leptons
$\ell_\alpha$) and $\gamma^{\alpha}_{\mathrm{W},\Delta L = 1}$.

\providecommand{\bysame}{\leavevmode\hbox to3em{\hrulefill}\thinspace}


\begin{thebibliography}{10}

\bibitem{Spergel:2006hy}
D.~N.~Spergel {\it et al.},
arXiv:astro-ph/0603449.

\bibitem{Kuzmin:1985mm}
  V.~A.~Kuzmin, V.~A.~Rubakov and M.~E.~Shaposhnikov,
  Phys.\ Lett.\ B {\bf 155} (1985) 36.

\bibitem{seesaw:I}
P.~Minkowski,
Phys.\ Lett.\ B {\bf 67} (1977) 421;
%
M.~Gell-Mann, P.~Ramond and R.~Slansky, in {\it Complex Spinors and
  Unified Theories} eds. P.~Van.~Nieuwenhuizen and D.~Z.~Freedman,
  {\it Supergravity} (North-Holland, Amsterdam, 1979), 
  p.315 [Print-80-0576 (CERN)];
%
T.~Yanagida, in {\it Proceedings of the Workshop on the Unified Theory
and the Baryon Number in the Universe}, eds. O.~Sawada and
A.~Sugamoto (KEK, Tsukuba, 1979), p.95;
%
S.~L.~Glashow, in {\it Quarks and Leptons}, eds. M.~L\'evy {\it et
al.} (Plenum Press, New York, 1980), p.687;
%
R.~N.~Mohapatra and G.~Senjanovi\'c,
Phys.\ Rev.\ Lett.\  {\bf 44} (1980) 912.

\bibitem{seesaw:II}
%
R.~Barbieri, D.~V.~Nanopolous, G.~Morchio and F.~Strocchi, 
Phys.\ Lett.\ B {\bf 90} (1980) 91;
%
R.~E.~Marshak and R.~N.~Mohapatra, {\it Invited talk given at Orbis
  Scientiae, Coral Gables, Fla., Jan. 14-17, 1980}, VPI-HEP-80/02;
%
T.~P.~Cheng and L.~F.~Li,
Phys.\ Rev.\ D {\bf 22} (1980) 2860;
%
M.~Magg and C.~Wetterich,
Phys.\ Lett.\ B {\bf 94} (1980) 61;
%
G. Lazarides, Q. Shafi and C. Wetterich,
Nucl. Phys. B181 (1981) 287; 
J.~Schechter and J.~W.~F.~Valle,
Phys.\ Rev.\ D {\bf 22} (1980) 2227;
%
R.~N.~Mohapatra and G.~Senjanovi\'c,
Phys.\ Rev.\ D {\bf 23} (1981) 165.

\bibitem{FY} M.~Fukugita and T.~Yanagida,
Phys.\ Lett.\ B {\bf 174} (1986) 45.

\bibitem{lept}
For a review containing an extensive list of references, see e.g.:
W.~Buchm\"uller, R.~D.~Peccei and T.~Yanagida,
  Ann.\ Rev.\ Nucl.\ Part.\ Sci.\  {\bf 55} (2005) 311
  [arXiv:hep-ph/0502169].

\bibitem{Casas:2001sr}
J.~A.~Casas and A.~Ibarra,
Nucl.\ Phys.\ B {\bf 618} (2001) 171
[arXiv:hep-ph/0103065].

\bibitem{barbieri}
R.~Barbieri, P.~Creminelli, A.~Strumia and N.~Tetradis,
Nucl.\ Phys.\ B {\bf 575} (2000) 61
[arXiv:hep-ph/9911315].

\bibitem{endoh} 
T.~Endoh, T.~Morozumi and Z.~h.~Xiong,
  Prog.\ Theor.\ Phys.\  {\bf 111} (2004) 123
  [arXiv:hep-ph/0308276]; 
  T.~Fujihara, S.~Kaneko, S.~Kang, D.~Kimura, T.~Morozumi and M.~Tanimoto,
  Phys.\ Rev.\ D {\bf 72} (2005) 016006
  [arXiv:hep-ph/0505076]. 

\bibitem{davidsonetal}
A.~Abada, S.~Davidson, F.~X.~Josse-Michaux, M.~Losada and A.~Riotto, 
JCAP {\bf 0604} (2006) 004
  [arXiv:hep-ph/0601083].

\bibitem{nardietal}
E.~Nardi, Y.~Nir, E.~Roulet and J.~Racker,
  JHEP {\bf 0601} (2006) 164
  [arXiv:hep-ph/0601084].
    
\bibitem{Abada:2006ea}
  A.~Abada, S.~Davidson, A.~Ibarra, F.~X.~Josse-Michaux, 
M.~Losada and A.~Riotto,
  arXiv:hep-ph/0605281.

\bibitem{Blanchet:2006be}
  S.~Blanchet and P.~Di Bari,
  arXiv:hep-ph/0607330.
 
\bibitem{Antusch:2006cw}
  S.~Antusch, S.~F.~King and A.~Riotto,
  arXiv:hep-ph/0609038.
  
\bibitem{Pascoli:2006ie}
  S.~Pascoli, S.~T.~Petcov and A.~Riotto,
  arXiv:hep-ph/0609125. 

\bibitem{Branco:2006ce}
  G.~C.~Branco, R.~G.~Felipe and F.~R.~Joaquim,
  arXiv:hep-ph/0609297.

\bibitem{Pascoli:2002qm}
  S.~Pascoli, S.~T.~Petcov and W.~Rodejohann,
  Phys.\ Lett.\ B {\bf 549} (2002) 177
  [arXiv:hep-ph/0209059];
  S.~Pascoli, S.~T.~Petcov and T.~Schwetz,
  Nucl.\ Phys.\ B {\bf 734} (2006) 24
  [arXiv:hep-ph/0505226].

\bibitem{LeptAndLFV}
For example, see:
   J.~R.~Ellis and M.~Raidal,
  Nucl.\ Phys.\ B {\bf 643} (2002) 229
  [arXiv:hep-ph/0206174]; 
  J.~R.~Ellis, M.~Raidal and T.~Yanagida,
  Phys.\ Lett.\ B {\bf 546} (2002) 228
  [arXiv:hep-ph/0206300];  
  S.~Pascoli, S.~T.~Petcov and C.~E.~Yaguna,
  Phys.\ Lett.\ B {\bf 564} (2003) 241
  [arXiv:hep-ph/0301095]; 
   S.~Kanemura, K.~Matsuda, T.~Ota, T.~Shindou, E.~Takasugi and K.~Tsumura,
  Phys.\ Rev.\ D {\bf 72} (2005) 093004
  [arXiv:hep-ph/0501228]; 
  S.~Kanemura, K.~Matsuda, T.~Ota, T.~Shindou, E.~Takasugi and K.~Tsumura,
  Phys.\ Rev.\ D {\bf 72} (2005) 055012
  [Erratum-ibid.\ D {\bf 72} (2005) 059904]
  [arXiv:hep-ph/0507264]; 
  S.~T.~Petcov, T.~Shindou and Y.~Takanishi,
  Nucl.\ Phys.\ B {\bf 738} (2006) 219
  [arXiv:hep-ph/0508243]; 
  S.~T.~Petcov, W.~Rodejohann, T.~Shindou and Y.~Takanishi,
  Nucl.\ Phys.\ B {\bf 739} (2006) 208
  [arXiv:hep-ph/0510404];
   F.~Deppisch, H.~Pas, A.~Redelbach and R.~Ruckl,
  Phys.\ Rev.\ D {\bf 73} (2006) 033004
  [arXiv:hep-ph/0511062].

\bibitem{Antusch:2006vw}
  S.~Antusch, E.~Arganda, M.~J.~Herrero and A.~M.~Teixeira,
  arXiv:hep-ph/0607263.

\bibitem{Giudice:2003jh}
G.~F.~Giudice, A.~Notari, M.~Raidal, A.~Riotto and A.~Strumia,
Nucl.\ Phys.\ B {\bf 685} (2004) 89
[arXiv:hep-ph/0310123].

\bibitem{Davidson:2002qv}
S.~Davidson and A.~Ibarra, 
Phys.\ Lett.\ B \textbf{535} (2002) 25 [arXiv:hep-ph/0202239].  

\bibitem{Guth:1980zm}
A.~H. Guth, 
Phys.\ Rev.\ D \textbf{23} (1981) 347;
%
A.~D. Linde, 
Phys.\ Lett.\ B \textbf{108} (1982) 389;
%
A.~Albrecht and P.~J. Steinhardt, 
Phys.\ Rev.\ Lett. \textbf{48} (1982) 1220. 
%
For a review containing an extensive list of references, 
see also: D.~H.~Lyth and A.~Riotto,
Phys.\ Rept.\  {\bf 314} (1999) 1 [arXiv:hep-ph/9807278].
  
\bibitem{Covi:1996wh}
  L.~Covi, E.~Roulet and F.~Vissani,
  Phys.\ Lett.\ B {\bf 384} (1996) 169
  [arXiv:hep-ph/9605319].

\bibitem{res}
M.~ Flanz, E.~A.~Paschos, U.~Sarkar and J.~Weiss, 
Phys.\ Lett.\ B {\bf 389} (1996) 693; 
A.~Pilaftsis, Phys.\ Rev.\ D {\bf 56} (1997) 5431.
For recent works where flavour effects are taken into account, see e.g.:  
A.~Pilaftsis and T.~E.~J.~Underwood,
Phys.\ Rev.\ D {\bf 72} (2005) 113001 [arXiv:hep-ph/0506107];
G.~C.~Branco, A.~J.~Buras, S.~Jager, S.~Uhlig and A.~Weiler,
arXiv:hep-ph/0609067. 

\bibitem{Vives:2005ra}
  O.~Vives,
  Phys.\ Rev.\ D {\bf 73} (2006) 073006
  [arXiv:hep-ph/0512160]. 
  
\bibitem{nonthermalLG} 
See e.g.:  
G.~Lazarides and Q.~Shafi,
Phys.\ Lett.\ B {\bf 258} (1991) 305;
H.~Murayama and T.~Yanagida, Phys.\ Lett.\ B {\bf 322} (1994) 349 
[arXiv:hep-ph/9310297];
K.~Hamaguchi, H.~Murayama and T.~Yanagida,  Phys.\ Rev.\ D {\bf 65} (2002) 
043512 [arXiv:hep-ph/0109030];
T.~Moroi and H.~Murayama,  Phys.\ Lett.\ B {\bf 553} (2003) 126
[arXiv:hep-ph/0211019]. 

\bibitem{SneutrinoInflation}
See e.g.:   
H.~Murayama, H.~Suzuki, T.~Yanagida and J.~Yokoyama,  
Phys.\ Rev.\ Lett.\ \textbf{70} (1993) 1912;
J.~R. Ellis, M.~Raidal and T.~Yanagida,  
Phys.\ Lett.\ B \textbf{581} (2004) 9 [arXiv:hep-ph/0303242];
S.~Antusch, M.~Bastero-Gil, S.~F.~King and Q.~Shafi,
Phys.\ Rev.\ D {\bf 71} (2005) 083519 [arXiv:hep-ph/0411298].
 
\bibitem{flatdir}
For recent discussions, see e.g.: 
  R.~Allahverdi and A.~Mazumdar,
  arXiv:hep-ph/0603244;
  K.~A.~Olive and M.~Peloso,
  arXiv:hep-ph/0608096;
  R.~Allahverdi and A.~Mazumdar,
  arXiv:hep-ph/0608296;
  R.~Allahverdi, K.~Enqvist, J.~Garcia-Bellido, A.~Jokinen and A.~Mazumdar,
  arXiv:hep-ph/0610134.
  
\bibitem{pedestrians}
W.~Buchm\"uller, P.~Di Bari and M.~Pl\"umacher,
Annals Phys.\  {\bf 315} (2005) 305
[arXiv:hep-ph/0401240].

\bibitem{Pilaftsis:2003gt}
  A.~Pilaftsis and T.~E.~J.~Underwood,
  Nucl.\ Phys.\ B {\bf 692} (2004) 303
  [arXiv:hep-ph/0309342].

\bibitem{Buchmuller:2001sr}
  W.~Buchm\"uller and M.~Pl\"umacher,
  Phys.\ Lett.\ B {\bf 511} (2001) 74
  [arXiv:hep-ph/0104189].

\bibitem{Antusch:2003kp}
S.~Antusch, J.~Kersten, M.~Lindner and M.~Ratz, 
Nucl.\ Phys.\ B \textbf{674} (2003) 401[arXiv:hep-ph/0305273].

\bibitem{Antusch:2005gp}
  S.~Antusch, J.~Kersten, M.~Lindner, M.~Ratz and M.~A.~Schmidt,
  JHEP {\bf 0503} (2005) 024
  [arXiv:hep-ph/0501272].

\bibitem{PDG} 
W.-M.~Yao {\it et al.}\ [Particle Data Group Collaboration], 
J.\ Phys.\ G {\bf 33} (2006) 1.

\bibitem{Lindner:2005kv}
  A recent discussion of possible differences between both treatments
  can be found in: 
  M.~Lindner and M.~M.~Muller,
  Phys.\ Rev.\ D {\bf 73} (2006) 125002
  [arXiv:hep-ph/0512147].
  
\bibitem{gravitinoproblem}
M.~Y.~Khlopov and A.~D.~Linde,  
Phys.\ Lett.\ B \textbf{138} (1984) 265;
J.~R.~Ellis, J.~E.~Kim and D.~V.~Nanopoulos,  
Phys.\ Lett.\ B \textbf{145} (1984) 181;
J.~R.~Ellis, D.~V.~Nanopoulos and S.~Sarkar,  
Nucl.\ Phys.\ B \textbf{259} (1985) 175;
T.~Moroi, H.~Murayama and M.~Yamaguchi,  
Phys.\ Lett.\ B \textbf{303} (1993) 289;
M.~Kawasaki, K.~Kohri and T.~Moroi,  
Phys.\ Lett.\ B {\bf 625} (2005) 7  [arXiv:astro-ph/0402490].
  
\bibitem{Kohri:2005wn}
 For a recent discussion, see:  
  K.~Kohri, T.~Moroi and A.~Yotsuyanagi,
  Phys.\ Rev.\ D {\bf 73} (2006) 123511
  [arXiv:hep-ph/0507245].

\bibitem{thermal}  
For thermal production of gravitinos, see e.g.:
  M.~Bolz, A.~Brandenburg and W.~Buchmuller,
  Nucl.\ Phys.\ B {\bf 606} (2001) 518
  [arXiv:hep-ph/0012052];
  J.~Pradler and F.~D.~Steffen,
  arXiv:hep-ph/0608344.
  
\bibitem{Roszkowski:2004jd}
  L.~Roszkowski, R.~Ruiz de Austri and K.~Y.~Choi,
  JHEP {\bf 0508} (2005) 080
  [arXiv:hep-ph/0408227]; 
  D.~G.~Cerde\~no, K.~Y.~Choi, K.~Jedamzik, 
  L.~Roszkowski and R.~Ruiz de Austri,
  JCAP {\bf 0606} (2006) 005
  [arXiv:hep-ph/0509275];
  see also: 
  F.~D.~Steffen,
   JCAP {\bf 0609} (2006) 001
   [arXiv:hep-ph/0605306].

\bibitem{hep-ph/0609142}
   W.~Buchm\"uller, L.~Covi, J.~Kersten and K.~Schmidt-Hoberg,
  arXiv:hep-ph/0609142.
  
\bibitem{Maltoni:2004ei}
For our numerical analysis, we have used the global fit results from: 
M.~Maltoni, T.~Schwetz, M.~A.~Tortola and J.~W.~F.~Valle,
  New J.\ Phys.\  {\bf 6} (2004) 122
  [arXiv:hep-ph/0405172v5].

\bibitem{Buchmuller:2003gz}
W.~Buchm{\"u}ller, P.~Di~Bari and M.~Pl{\"u}macher, 
  Nucl.\ Phys.\ B \textbf{665} (2003) 445
  [arXiv:hep-ph/0302092].
  
\bibitem{Antusch:2004xy}
  S.~Antusch and S.~F.~King,
  Phys.\ Lett.\ B {\bf 597} (2004) 199
  [arXiv:hep-ph/0405093].

\bibitem{Antusch:2005tu}
  S.~Antusch and S.~F.~King,
  JHEP {\bf 0601} (2006) 117
  [arXiv:hep-ph/0507333];
  Phys.\ Lett.\ B {\bf 591} (2004) 104
  [arXiv:hep-ph/0403053]; 
  Nucl.\ Phys.\ B {\bf 705} (2005) 239
  [arXiv:hep-ph/0402121].

\bibitem{Aalseth:2004hb}
  For a summary and references, see e.g.: C.~Aalseth {\it et al.} 
  [Part of the APS Neutrino Study],
  arXiv:hep-ph/0412300.

\bibitem{King:2006hn}
  S.~F.~King,
  arXiv:hep-ph/0610239.
  
\end{thebibliography}
\end{document}